\documentclass[journal=jctcce,manuscript=article]{achemso}

\usepackage[version=3]{mhchem} 
\usepackage{graphicx}
\usepackage{braket}
\usepackage{multirow}
\usepackage{amsfonts}
\usepackage{threeparttable}
\usepackage{siunitx}
\usepackage{xcolor}
\usepackage{appendix}
\usepackage{natbib}
\usepackage{amsmath}

\newcommand{\ehs}[2]{\ket{#1}\bra{#2}}
\newcommand{\elsket}[2]{\ket{\hat{#1}_{#2}}}
\newcommand{\elsbra}[2]{\bra{\hat{#1}_{#2}}}
\newcommand{\proj}[2]{\ket{#1}\bra{#2}}
\newcommand{\tr}[1]{\mathrm{Tr}#1}
\newcommand{\bs}[1]{\boldsymbol{#1}}
\newcommand{\dhat}[1]{\hat{\hat{#1}}}

\author{Tzu Yu Wang}
\email{joeywang05@gmail.com}
\affiliation[uOttawa]
{Department of Chemistry and Biomolecular Sciences, University of Ottawa, Ottawa, Canada}
\author{Michael S. Schuurman}
\email{michael.schuurman@uottawa.ca}
\affiliation[NRC]
{National Research Council Canada, 100 Sussex Dr., Ottawa, Canada, K1A 0R6}
\alsoaffiliation[uOttawa]
{Department of Chemistry and Biomolecular Sciences, University of Ottawa, Ottawa, Canada}
\author{Simon P. Neville}
\email{simon.neville@nrc-cnrc.gc.ca}
\affiliation[NRC]
{National Research Council Canada, 100 Sussex Dr., Ottawa, Canada, K1A 0R6}
\title[OSD]
  {Operator Entanglement in Quantum Dynamics Simulations: Formalism and Analysis Tools}

\abbreviations{OSD}

\begin{document}

\begin{abstract}
We review the framework of operator Hilbert space and introduce the one- and two-particle super reduced density matrices (1-SRDMs and 2-SRDMs), as well as the super mutual information (SMI). The eigenvectors of the 1-SRDMs define what we term natural single particle operator bases, and provide a way to compress vibrational and vibronic Hamiltonians with controlled error. The SMI is defined from the operator entanglement entropy of the 1-SRDMs and 2-SRDMs, and captures the correlation between operators acting on different one-mode subspaces, which may be used to reveal and quantify both direct and indirect couplings that might otherwise be difficult to extract. Efficient numerical approaches for the calculation of SRDMs and the SMI are developed and applied to a set of prototypical vibrational and vibronic Hamiltonians, as well as approximations to the corresponding time-evolution operators. Through this, we demonstrate that: (i) commonly used vibronic Hamiltonians are amenable to extremely high levels of compression without compromising accuracy, and; (ii) SMI analysis can be used to systematically and quantitatively reveal both direct and indirect couplings that might otherwise be difficult to extract, including indirect couplings of vibrational modes via intermediary electronic-vibrational interactions.

\end{abstract}
\maketitle

\section{Introduction}

The inherently high dimensionality of molecular systems presents a fundamental challenge in quantum dynamics simulations: representing the quantum state numerically exactly in a tensor product basis is generally infeasible due to the exponential scaling of the basis size with the number of degrees of freedom. A widely used strategy to avoid exponential scaling is based on the observation that many of the physically relevant quantum many-body states occupy only a small ``corner'' of the Hilbert space\cite{PhysRevLett.106.170501}. This insight motivates the use of variational classes of wavefunctions with non-exponential scaling, chosen to efficiently capture the essential features of the state of interest.

Tensor network states (TNS) have emerged as powerful variational classes characterized by their ability to capture and compactly represent correlation in many-body systems \cite{Orus2019, eisert2013entanglementtensornetworkstates, ORUS2014117, bridgeman2017hand}. Although the formalism of TNSs have matured and unified in recent years\cite{Ran2020chp1, Larsson17072024,SCHOLLWOCK201196,Evenbly2011}, its conceptual foundations appeared independently across disciplines – for instance, the multi-layer multi-configurational time-dependent Hartree (ML-MCTDH) in chemical physics\cite{10.1063/1.1580111,MEYER199073,10.1063/1.2902982,10.1063/1.3535541}, the density matrix renormalization group (DMRG) in condensed matter physics\cite{PhysRevLett.69.2863,whiteprb_dmrg}, and graph states in quantum computing\cite{hein2006entanglementgraphstatesapplications,PhysRevA.69.062311}. The underlying challenge across all these disciplines is that, given a basis, the coefficient tensor of a quantum many-body state scales exponentially with the number of degrees of freedom. TNS addresses this by approximating the high-dimensional tensor through a decomposition into a network of contractions of low-dimensional tensors. Diagrammatically, TNSs can be represented as graphs, where each physical degrees of freedom corresponds to a node with open (uncontracted) indices, and contracted edges encode correlation\cite{Ran2020chp2,eisert2013entanglementtensornetworkstates,ORUS2014117,Evenbly2011}.

Within the TNS frameowrk, the underlying \textit{ansatze} used in ML-MCTDH and DMRG correspond, respectively, to a tree-tensor network state (TTNS) and matrix product state (MPS). MPS in particular is a simple but powerful variational class that, once a basis is chosen, corresponds to factorizing the $d$-dimensional coefficient tensor into a contraction of $d$ three-dimensional tensors\cite{McCulloch_2007,SCHOLLWOCK201196}. While an MPS is a limiting case of the more general TTNS used in ML-MCTDH, MPS is often adequate for studying vibronic problems with a much simpler parametrization and equations of motion using the time-dependent variational principle\cite{https://doi.org/10.1002/wcms.1614,doi:10.1021/acs.jctc.4c00751,doi:10.1021/acs.jctc.9b00301}. The dimension of the contracted edges in a TNS, which encode correlation, can be understood and interpreted physically through notions of entanglement entropy. Originally devised by White\cite{PhysRevLett.69.2863}, the feasibility of approximating the ground state of one-dimensional spin-chain Hamiltonian with DMRG is directly related to the entanglement entropy across bi-partitions\cite{SCHOLLWOCK201196,McCulloch_2007,Catarina2023}. In the TNS framework, this signifies that the dimensions and computational efficiency of the MPS approximation can be inferred from entanglement entropy measures. Understanding how the entanglement entropy grows with system size thus informs us whether MPS is a suitable variational class for the problem under consideration. For quantum dynamical problems, beyond its scaling with system size, this consideration further translates to how the entanglement entropy grows in time. Entanglement is therefore an important quantity that connects the choice of the variational class to the problem at hand.

Entanglement entropy measures can be extended beyond quantum \textit{states} to elements in any many-body Hilbert space, notably to the space of linear \textit{operators}. The early works on operator entanglement entropy (OEE) and the entangling capability of unitary operators were done in the context of quantum information theory. Operator entanglement was first introduced by Zanardi \textit{et al.}\cite{PhysRevA.63.040304}, by considering the space of operators forming a Hilbert space and defining the operator linear entanglement entropy $1-\tr{((\hat{\rho}^{U})^2)}$, where $\hat{\rho}^U$ is the density operator of the vectorized unitary operator $\hat{U}$. They showed that the linear operator entanglement entropy is related to the entangling power\cite{PhysRevA.62.030301}, defined as the average entanglement created over all possible product states\cite{PhysRevA.63.040304}. Nielsen \textit{et al.} used the operator entanglement entropy defined as the Von Neuman entanglement entropy and showed that the OEE of a unitary operator is a lower bound to the maximum entanglement a unitary can create between two initially unentangled systems with the use of arbitrary ancillas\cite{PhysRevA.67.052301}. A summary of other results where the operator is analyzed can be found in Nielsen \textit{et al.} \cite{PhysRevA.67.052301}. This set of results connects the entanglement generated in a state by the action of a unitary operator to the entanglement measure defined at the operator level. Specifically, in the context of quantum dynamics, probing the entanglement entropy of the time-evolution operator is related to how much entanglement can be created during the dynamics. 

As far as we are aware, this is a topic that remains unexplored within the molecular quantum dynamics community. In the physics community, however, OEE has been used to characterize the out-of-equilibrium quantum many-body system dynamics\cite{PhysRevB.98.174304, PhysRevLett.122.250603}. The time dependence of the operator entanglement entropy for integrable and non-integrable systems \cite{PhysRevB.98.174304, PhysRevLett.122.250603} have been studied and proposed as a method for characterizing distinct dynamic regimes such as chaotic and many-body localized systems\cite{PhysRevB.98.174304, PhysRevLett.122.250603, PhysRevB.95.094206}. Beyond analyzing the OEE of the Hamiltonian or different unitary operators, an apparent case where it is crucial to carry out the entanglement analysis at the operator level is when studying mixed state or open quantum dynamics, where the time evolution must be carried out on the density operator.

In this work, we review the theory needed to understand and utilize OEE-based analysis tools, including super reduced density matrices (SRDMs) and the super mutual information (SMI). For a selection of prototypical vibrational and vibronic Hamiltonians, we show that the representation in terms of the eigenvectors of the one-body SRDMs allows for a massive compression of the Hamiltonian without compromising accuracy. We also demonstrate how the SMI can be used to reveal and quantify couplings between modes at the operator level, including the indirect coupling of vibrational modes mediated via direct coupling to the electronic degree of freedom.

\section{Formalism for States}



\subsection{Hilbert Subspace Truncation}\label{subsec:H_subspace_trunc}

To introduce super-operators with clarity, it would benefit to recall the analogous quantities for states in a Hilbert space $\mathcal{H}$, including the role of the reduced density in the truncation of quantum states and the construction of entanglement entropy and the mutual information matrix. The key insight is that once the underlying Hilbert space $\mathcal{H}$ for a set of operators is defined, all the standard ideas used in the Hilbert space of quantum states then rigorously apply. This includes the super reduced density matrices and the super mutual information, which will be detailed in the next section. The following derivation is the foundation for the density matrix renormalization group (DMRG) method\cite{PhysRevLett.69.2863}, and follows mostly from Catarina et al \cite{Catarina2023}.

For a $d$-body system, the state $\ket{\Psi}\in\mathcal{H}$, where $\mathcal{H}=\mathcal{H}^{(1)}\otimes \mathcal{H}^{(2)}\otimes\cdots\otimes \mathcal{H}^{(d)}$, and $\{\ket{i_\kappa}\}\in \mathcal{H}^{(\kappa)}$ is a complete basis for the local Hilbert space of particle $\kappa$ of dimension $N_\kappa$, is given by

\begin{equation}\label{eq:general_N_body}
    \ket{\Psi}=\sum_{i_1=1}^{N_1}\cdots\sum_{i_d=1}^{N_d}C_{i_1,\dots, i_d}\ket{i_1}\otimes\cdots\otimes\ket{i_d},
\end{equation}

\noindent with the position representation wavefunction given by 

\begin{equation}\label{eq:general_N_body_Q_rep}
    \begin{aligned}\braket{\boldsymbol{Q}|\Psi}&=\ket{\Psi(\boldsymbol{Q})}\equiv\sum_{i_1=1}^{N_1}\cdots\sum_{i_d=1}^{N_d}C_{i_1,\dots, i_d}\braket{Q_1|i_1}\otimes\cdots\otimes\braket{Q_d|i_d} \\
    &=\sum_{i_1=1}^{N_1}\cdots\sum_{i_d=1}^{N_d}C_{i_1,\dots, i_d}\ket{i_1(Q_1)}\otimes\cdots\otimes\ket{i_d(Q_d)}
    \end{aligned}
\end{equation}

We consider an exact bi-partitioning of the system into two subspaces, $A$ and $B$, with associated Hilbert spaces 

\begin{subequations}
\label{eq:system}
    \begin{align}
        &\mathcal{H}^{(X)} =\bigotimes_{\kappa\in X}\mathcal{H}^{(\kappa)}, ~~ X=A,B\\
        &N_{X} = \mathrm{dim}(\mathcal{H}^{(X)}) = \prod_{\kappa\in X}\mathrm{dim}(\mathcal{H}^{(\kappa)}), ~~ X=A,B
    \end{align}
\end{subequations}

\noindent
The state $|\Psi\rangle$ may then be written as

\begin{equation}\label{eq:general_N_body_bipartite}
    \ket{\Psi}=\sum_{i_A=1}^{N_A}\sum_{i_B=1}^{N_B}C_{i_A,i_B}\ket{i_A}\otimes\ket{i_B},
\end{equation}

\noindent where $\ket{i_{X}}=\bigotimes_{\kappa \in X}\ket{i_\kappa},~X=A,B$ are the bi-partite bases for the Hilbert subspaces $A$ and $B$. It is clear from Eq.~\eqref{eq:general_N_body_Q_rep} that the coefficient tensor $C_{i_1,\dots,i_d}$ scales exponentially with the number of degrees of freedom. In many variational approaches, including that of the DMRG method\cite{PhysRevLett.69.2863}, we seek a variational state

\begin{equation}\label{eq:reduced_N_body_bipartite}
    \ket{\tilde{\Psi}}=\sum_{\alpha_A=1}^{D_A}\sum_{i_B=1}^{N_B}\tilde{C}_{\alpha_A,i_B}\ket{\alpha_A}\otimes\ket{i_B},
\end{equation}

\noindent that minimizes the $L_2$ norm $||\ket{\Psi}-\ket{\tilde{\Psi}}||_2$, where ideally $D_A \ll N_A$. Substituting the set of coefficients that satisfies $\frac{\partial}{\partial C_{\alpha_A,i_B}^*}||\ket{\Psi}-\ket{\tilde{\Psi}}||_2=0$ into Eq.~\eqref{eq:reduced_N_body_bipartite} then leads to

\begin{equation}\label{eq:max_trunc_condition}
    \left|\left|\ket{\Psi}-\ket{\tilde{\Psi}}\right|\right|_2 = 1 -\sum_{\alpha_A=1}^{D_A}\braket{\alpha_A|\hat{\rho}_A|\alpha_A},
\end{equation}

\noindent where the state $\ket{\Psi}$ is assumed to be normalized, and in Eq.~\eqref{eq:max_trunc_condition} the reduced density operator

\begin{equation}
    \hat{\rho}_A=\mathrm{Tr}_B(\hat{\rho})=\sum_{i_A=1}^{N_A}\sum_{i_B=1}^{N_B}\sum_{i_{A}'=1}^{N_A}C_{i_A,i_B}^*C_{i'_A,i_B}\ket{i_A}\bra{i'_A},
\end{equation}

\noindent naturally emerges. It is clear from Eq.~\eqref{eq:max_trunc_condition} that minimizing the error arising from the truncation amounts to choosing the basis $\ket{\alpha_A}$ that maximizes the expectation value of $\hat{\rho}_A$. It can be shown using a corollary of the Schur-Horn theorem that the states that maximizes this value are precisely the eigenvectors of the operator $\hat{\rho}_A$\cite{Catarina2023}. Eq.~\eqref{eq:max_trunc_condition} then evaluates to a sum over the eigenvalues of $\hat{\rho}_A$,

\begin{equation}
    \left|\left|\ket{\Psi}-\ket{\tilde{\Psi}}\right|\right|_2 = 1-\sum_{i_A=1}^{N_A}\lambda_{i_A}=\sum_{i_A=D_A}^{N_A}\lambda_{i_A}, 
\end{equation}

\noindent
and the accuracy of the truncated variational wavefunction thus depends on the spectrum decay of $\hat\rho_A$, where the eigenvalues are ordered by descending order $\lambda_1\geq\lambda_2\geq\cdots\geq\lambda_d$, and for a normalized state $\sum_{i_A=1}^{N_A}\lambda_{i_A}=1$.  By retaining only the first $D\leq N_A$ eigenstates in the construction of $\ket{\tilde{\Psi}}$, the error between the approximate and exact state is thus the sum over the discarded eiganvalues given. An important result from this review is that the reduced density naturally emerges in the approximate truncation of quantum states, and the eigenstates of the reduced density are the optimal truncation basis for a system with two degrees of freedom. In general, for a many-body system the equality is replaced with an upper bound as\cite{PhysRevB.73.094423}

\begin{equation}\label{eq:accuracy_from_rdm_3}
    ||\ket{\Psi}-\ket{\tilde{\Psi}}||_2\leq \sum_{\kappa=1}^{d}\sum_{i_\kappa=D_\kappa+1}^{N_\kappa}\lambda_{i_\kappa}
\end{equation}

Importantly, this analysis only relies on the existence and subsequent properties of a Hilbert space $\mathcal{H}$, and as such one may expect this procedure to hold for any other valid Hilbert space that does not have quantum states as its elements. This is the main premise for the operator-based analysis that is introduced in later sections.

\subsection{Mutual Information}\label{subsec:mi}

As seen from Eq.~\eqref{eq:accuracy_from_rdm_3}, the rate of decay of the spectrum of $\hat{\rho}_A$ determines the compressibility and approximability of the corresponding state $|\Psi\rangle$. A central concept that quantifies this idea is the Von-Neumann entanglement entropy (vNEE), which considers the bi-partite entanglement between two subspaces of a pure state. While we only consider states that are pure in the full Hilbert space, bi-partitions can still result in a mixed states that give rise to entangled dynamics between subspaces. The definition of the bi-partite vNEE for a pure state is

\begin{equation}\label{eq:vnee}
    S_A\equiv S(\hat{\rho}_A)\equiv-\mathrm{Tr}(\hat{\rho}_A \mathrm{log}\hat{\rho}_A)=-\sum_{i_A=1}^{N_A}\lambda_{i_A}\mathrm{log}\lambda_{i_A},
\end{equation}

\noindent where the same notation for the reduced density matrix (RDM) $\hat{\rho}_A$ and its eigenvalues $\lambda_{i_A}$ from above is used. For molecular systems, the degrees of freedom contained in $A$ and $B$ would correspond to the vibrational modes plus the electronic degree of freedom, and $A\cup B$ is the complete set of vibrational and electronic degrees of freedom. It can be shown that the scaling of the vNEE sets a lower bound and the approximate size of the truncated dimension $D_A$. More specifically, the dimension of the the truncated Hilbert space is upper bounded in $S_A$ as $D_A\geq e^{S_A}$ \cite{Catarina2023, RevModPhys.82.277}. The study of how the vNEE $S_A$ scales with either the system size or with time for different states and different classes of Hamiltonian thus informs us of their approximability. This is a non-trivial problem and is not discussed here, but we can use it to motivate the use of the vNEE as a proxy for the feasibility of approximate states.

It is natural then to extend the vNEE to map out the entanglement structure inherent to the wavefunction between all bi-partitions. One approach is to examine the mutual information. From the subadditivity property of the vNEEs, the joint vNEE of $A$ and $B$, $S_{AB}$, is related to the single particle vNEEs as $S_{AB}\leq S_A+S_B$, with the equality holding only when $A$ and $B$ are unentangled\cite{Carlen2009TRACEIA}. This inequality allow us to determine the entanglement between $A$ and $B$ by considering the difference $S_A+S_B-S_{A,B}\geq0$. The mutual information (MI) matrix $I$ is then formed by constructing the joint vNEE between all pairs of degrees of freedom, with elements given by\cite{RISSLER2006519}:

\begin{equation}\label{eq:mi}
    I_{ij}=\frac{1}{2}(S_i+S_j-S_{ij})(1-\delta_{ij})\geq0\quad,
\end{equation}

\noindent where the diagonal portion of the vNEE does not contribute and is set to zero. The mutual information matrix maps out the entanglement between pairs of Hilbert subspaces $\mathcal{H}^{(i)}$ and $\mathcal{H}^{(j)}$ which, for molecular systems, provides a measure of the vibrational-vibrational and vibrational-electronic entanglement.


\section{Formalism for Operators}
\subsection{Hilbert Schmidt Operators}
The ideas of Hilbert space truncation, vNEE, and mutual information analysis can be readily extended to operators by introducing Hilbert–Schmidt operators and the associated operator Hilbert space.

\textbf{Definition:} 
A compact linear operator $\hat{A}$ on a Hilbert space $\mathcal{H}$ is a Hilbert-Schmidt operator on $\mathcal{H}$ if it satisfies\cite{Gohberg1990}

\begin{equation}\label{eq:hs_operator_condition}
    \begin{aligned}
        \sum_i ||\hat{A}\phi_i||^2&<\infty\\
    \end{aligned}
\end{equation}

\noindent for any orthonormal basis $\{\phi_i\}$ of $\mathcal{H}$. Note that if Eq.~\eqref{eq:hs_operator_condition} is satisfied, then it can be easily shown that 

\begin{equation}\label{eq:hs_operator_condition_2}
    \sum_{i,j}|\braket{\hat{A}\phi_i,\phi_j}|^2<\infty
\end{equation}

\noindent is also satisfied\cite{Gohberg1990}. The set of all Hilbert-Schmidt operators on a Hilbert space $\mathcal{H}$ forms a Hilbert space\cite{Gohberg1990,Sołtan2018}, denoted as $B_2(\mathcal{H})$, with respect to the Hilbert-Schmidt inner product:

\begin{equation}\label{eq:Hilbert-schmidt inner product}
    \braket{\hat{A},\hat{B}}_{HS}=\mathrm{Tr}(\hat{A}^{\dagger}\hat{B}),
\end{equation}

\noindent with the norm induced by the inner product denoted as $||\cdot||_{HS}$.

A complete basis for $B_2(\mathcal{H})$ is the set of $\{e_i\otimes e^j\}$, where $\{e_i\}$ and $\{e^j\}$ is a basis for $\mathcal{H}$ and its dual, respectively. The simplest complete basis in $B_2(\mathcal{H})$ is therefore the set $\ket{i}\bra{j}$, where $\{\ket{i}\}$ is complete basis set for $\mathcal{H}$. Recognizing that $B_2(\mathcal{H})$ is a Hilbert space with respect to the Hilbert-Schmidt inner product, we can then use the computational quantities in an analogous way for operators as that of states. Notably, the completeness of, for example, the basis 

\begin{equation}
    \{\hat{E}_I\} = \{\ket{i}\bra{j} ~|~ \ket{i}, \ket{j} \in \mathcal{H}\}
\end{equation}

\noindent
means any operator in $\hat{O} \in B_2(\mathcal{H})$ can be expressed as 

\begin{equation}
    \hat{O}=\sum_I\braket{\hat{O},\hat{E}_I}_{HS}\hat{E}_I.    
\end{equation}

Just as an operator is a linear map from one state to another state in the Hilbert space $\mathcal{H}$, $\hat{A}:\mathcal{H}\rightarrow\mathcal{H}$, linear maps that map an operator to another operator in $B_2(\mathcal{H})$ are called super-operators $\hat{\hat{A}}:B_2(\mathcal{H})\rightarrow B_2(\mathcal{H})$. Recognizing that a set of Hilbert-Schmidt operators form a Hilbert space $B_2(\mathcal{H})$, we can then rigorously define and apply many of the concepts usually associated for $\ket{\Psi}\in\mathcal{H}$ to $\hat{A}\in B_2(\mathcal{H})$. This is the key to being able to transfer the tools of state truncation and entanglement entropy analysis over to operators. First, however, we have to ensure that an operator of interest is indeed Hilbert-Schmidt. 

A corollary of Eq.~\eqref{eq:hs_operator_condition_2} is that if an operator has bounded singular values $s_j$, i.e., $\sum_j s_j^2<\infty$, then it is a Hilbert-Schmidt operator\cite{Gohberg1990}. This implies that all the operators in numerical variational approaches represented using finite bases are Hilbert-Schmidt, as they correspond to a projection onto the Hilbert subspace spanned by the set of chosen single particle basis. That is, any operator $\hat{O}$ can be represented exactly as 

\begin{equation}
\begin{aligned}
    \hat{1}\hat{O}\hat{1} &= (\hat{P}+\hat{Q})\hat{O}(\hat{P}+\hat{Q}) \\
    &= \hat{P}\hat{O}\hat{P}+\hat{P}\hat{O}\hat{Q}+\hat{Q}\hat{O}\hat{P}+\hat{Q}\hat{O}\hat{Q},
\end{aligned}
\end{equation}

\noindent
where $\hat{P}$ is the projector onto the chosen finite basis, and $\hat{Q}$ its complement. In numerical approaches, we discard all but the $\hat{P}\hat{O}\hat{P}$ term, which results in finite dimensional operators. More generally, it can be shown that\cite{Sołtan2018}

\begin{equation}\label{eq:operator_class_relation}
    {F}(\mathcal{H})\subset B_1(\mathcal{H})\subset B_2(\mathcal{H})\subset B_0(\mathcal{H}) ,
\end{equation}

\noindent where $B_0(\mathcal{H})$ is the set of compact operators on $\mathcal{H}$, $B_1(\mathcal{H})$ is the trace class, and $F(\mathcal{H})$ is the finite dimensional class. As a result, finite-dimensional representations of operators are Hilbert-Schmidt. This is crucial as even formally unbounded operators, such as the kinetic energy operator, have Hilbert-Schmidt representations in finite-dimensional numerical simulations. Thus, Hilbert space truncation and entanglement entropy based analysis tools can be applied to any operator encountered in practical quantum dynamics simulations.

To this end, in the next section, we define the relevant quantities pertaining to this: the super-reduced density matrix (SRDM), the natural single particle operator (NSPO) basis, and the super mutual information (SMI) matrix.

\subsection{Super Reduced Density Matrices}\label{sec:srdms}

By establishing the operator Hilbert space $B_2(\mathcal{H})$ and recognizing that any finite dimensional operator is an element in it, we can form quantities analogous to those on $\mathcal{H}$, such as the reduced density operator, with a ``super" prefix added to distinguish them from their usual state Hilbert space counterparts. Importantly, these quantities can be rigorously defined and justified. Utilizing the fact that $\hat{O}\in B_2(\mathcal{H})$ for any operator $\hat{O}$ used in practical finite-dimensional simulations, we can form an analogous super density operator for $\hat{O}$ as 

\begin{equation}
    \hat{\hat{R}}=\elsket{O}{}\elsbra{O}{},     
\end{equation}

\noindent
where the ket and bra notation of an operator is used to denote that $\elsket{O}{}\in B_2(\mathcal{H})$ and, similarly, the bra as the linear form of $B_2(\mathcal{H})$. Starting with the general exact bi-partite form of the operator 

\begin{equation}\label{eq:exact_H_bipartite_form}
    \begin{aligned}
        \hat{O}&= \sum_{i_Ai_B}\sum_{j_Aj_B} O_{i_A,i_B,j_A,j_B}\ehs{i_A}{j_A}\otimes\ehs{i_B}{j_B} \\
        &\equiv \sum_{I_A}\sum_{I_B}O_{I_A,I_B}\elsket{I}{A}\otimes\elsket{I}{B},
    \end{aligned}
\end{equation}

\noindent
with

\begin{equation}
    \elsket{I}{X}\equiv\ehs{i_{X}}{j_{X}}\in B_2(\mathcal{H}_{X}), ~~ X=A,B,
\end{equation}

\noindent
the super reduced density matrix (SRDM) over the subsystem $A$ then simply follows algebraically as

\begin{equation}\label{eq:1srdo_exact}
    \begin{aligned}
        \dhat{R}_A&=\mathrm{Tr}_{B}(\dhat{R}) = \sum_{L_B}\elsbra{L}{B}\hat{O}\rangle\langle\hat{O}\elsket{L}{B} \\
        &= \sum_{L_B}\sum_{I_A}\sum_{I'_A}O_{I_A,L_B}O^*_{I'_A,L_B}\elsket{I}{A}\elsbra{I'}{A} ,
    \end{aligned}
\end{equation}

\noindent where the inner product taken above is the Hilbert-Schmidt inner product defined in Eq.~\eqref{eq:Hilbert-schmidt inner product}, and the orthonormality of the Hilbert Schmidt basis was used: $\braket{\hat{I}_A|\hat{J}_A}=\delta_{IJ}$.

A clarification regarding the trace of different spaces should be noted. When taking the trace over an operator $\hat{O}: \mathcal{H}\rightarrow\mathcal{H}$, it is with respect to a basis $\{\ket{i}|\ \ket{i}\in\mathcal{H}\}$, including in the definition of the Hilbert-Schmidt inner product. Likewise, when taking the trace over a super-operator $\dhat{O}: B_2(\mathcal{H})\rightarrow B_2(\mathcal{H})$, it is with respect to an operator basis $\{\elsket{I}{} | \ \elsket{I}{}\in B_{2}(\mathcal{H})\}$, which is the case in Eq.~\eqref{eq:1srdo_exact}. For many calculations, working directly in the basis of $B_2(\mathcal{H})$ is often more convenient. Of course, in the latter case the inner products can always be re-expressed in the basis of $\mathcal{H}$, and the traces and inner products are taken in the usual sense.

For a general $d$-body system, the matrix element of the $\kappa^{\text{th}}$ one-particle SRDM (1-SRDM) is given by

\begin{equation}\label{eq:1srdm}
    \left(R^{(\kappa)}\right)_{ij,i'j'}=\sum_{I^\kappa}\sum_{J^\kappa}O_{I^\kappa_i,J^\kappa_j}O^*_{I^\kappa_{i'},J^\kappa_{j'}}
\end{equation}

\noindent
where the composite indices $L^{\kappa}\equiv l_1,\dots, l_{\kappa-1},l_{\kappa+1},\dots,l_d$ and $L^\kappa_i\equiv l_1,\dots, l_{\kappa-1},i,l_{\kappa+1},\dots,l_d$ is used. The 1-SRDM is thus a four-dimensional tensor $\boldsymbol{R}^{(\kappa)}\in\mathbb{C}^{N_\kappa\times N_\kappa\times N_\kappa\times N_\kappa}$.

The two-particle SRDM (2-SRDM) is obtained analogously by taking the subsystem $A$ to be comprised of two degrees of freedom, $\kappa$ and $\lambda$. We denote this by $\bs{R}^{(\kappa,\lambda)}$, and it takes the form

\begin{equation}\label{eq:2srdm}
    \left(R^{(\kappa,\lambda)}\right)_{ijkl,i'j'k'l'}=\sum_{I^{\kappa\lambda}}\sum_{J^{\kappa\lambda}}O_{I^{\kappa\lambda}_{ij},J^{\kappa\lambda}_{kl}}O^*_{I^{\kappa\lambda}_{i'j'},J^{\kappa\lambda}_{k'l'}}  ,
\end{equation}

\noindent 
where $L^{\kappa\lambda}\equiv l_1,\dots, l_{\kappa-1},l_{\kappa+1},\dots, l_{\lambda-1},l_{\lambda+1},l_d$ and $L_{ij}^{\kappa\lambda}\equiv l_1,\dots, l_{\kappa-1},i,l_{\kappa+1},\dots, l_{\lambda-1},j,l_{\lambda+1},l_d$.

\subsection{Natural Single Particle Operators}

In analogy with the natural orbitals encountered in quantum chemistry, which are the eigenvectors of the one-electron reduced density matrix, we term the eigenvectors of the 1-SRDM the natural single particle operators (NSPOs). That is, let $R^{(\kappa)}_{IJ}\in\mathbb{R}^{N_\kappa^2\times N_\kappa^2}$ be the matricized representation of the 1-SRDM in any basis $\{\elsket{I}{}\}$, then given its eigendecomposition 

\begin{equation}
    \boldsymbol{R}^{(\kappa)}\boldsymbol{U}^{(\kappa)}=\boldsymbol{U}^{(\kappa)}\boldsymbol{\Lambda}^{(\kappa)},    
\end{equation}

\noindent
the NSPO basis $\{ \elsket{\Upsilon}{\kappa} \}$ is given by

\begin{equation}
    \elsket{\Upsilon}{\kappa} = \sum_{I_{\kappa}=1}^{N_{\kappa}^{2}} U_{\Upsilon_{\kappa},I_{\kappa}}^{(\kappa)} \elsket{I}{\kappa}
\end{equation}

\subsection{Operator compression with NSPOs}\label{sec:op_compression_nspo}

As seen from Section~\ref{subsec:H_subspace_trunc}, the error of an approximate quantum state in a Hilbert space is bounded when truncated with the eigenvectors of the 1-RDM. Importantly, the requirement in the derivation of the error bound is only the existence of the inner product and induced norm of the underlying Hilbert space. Therefore, in close analogy with the truncation of states, the truncation of an operator via the NSPO basis in $B_2(\mathcal{H})$ is bounded with respect to the Hilbert-Schmidt norm. 

Retaining the $M_{\kappa} < N_{\kappa}^{2}$ NSPOs $\elsket{\Upsilon}{\kappa}$ with the largest eigenvalues $\Lambda_{J}^{(\kappa)}$ for each degree of freedom $\kappa$, and taking the eigenvalues to be in descending order, the operator $\hat{O}$ can be approximated as

\begin{equation}\label{eq:O_nspo_general_form}
    \begin{aligned}
        \hat{O} &= \sum_{I_{1}=1}^{N_{1}^{2}} \cdots \sum_{I_{d}=1}^{N_{2}^{2}} O_{I_{1},\dots,I_{d}} 
        \elsket{I}{1} \otimes \cdots \elsket{I}{d} \\
        &\approx \sum_{\Upsilon_1=1}^{M_1}\cdots\sum_{\Upsilon_d=1}^{M_d} \tilde{O}_{\Upsilon_1,\dots,\Upsilon_d}\elsket{\Upsilon}{1}\otimes\cdots\otimes\elsket{\Upsilon}{d},
    \end{aligned}
\end{equation}

\noindent
where

 \begin{equation}\label{eq:1srdm_truncated_coeff_tensor}
    \tilde{O}_{J_1,\dots,J_d} =\sum_{I_1=1}^{N_{1}^{2}} \cdots \sum_{I_d=1}^{N_{d}^{2}} O_{I_1,\dots,I_d}U^{\dagger}_{I_1 J_1}\cdots U^{\dagger}_{I_d J_d}
 \end{equation}

\noindent
We now comment on the error introduced by the expansion of the operator $\hat{O}$ in the truncated NSPO basis (Eqs.~\ref{eq:O_nspo_general_form}~and~\ref{eq:1srdm_truncated_coeff_tensor}). It can be shown that this truncation corresponds to a truncated higher-order singular value decomposition (HOSVD) of the original coefficient tensor $\bs{O}$ in the basis $\{ \elsket{I}{} \}$. It then follows that the error is bounded by the sum of the discarded eigenvalues\cite{T-HOSVD_2012},

\begin{equation}\label{eq:matrix_operator_truncation_bound}
    ||\bs{O}-\tilde{\bs{O}}||^2_{F} \leq \sum_{\kappa=1}^d \sum_{I=M_{\kappa}+1}^{N_{\kappa}^{2}} \Lambda_{I}^{(\kappa)},
\end{equation}

\noindent
where $||\cdot||_F$ denotes the Frobenius norm. In a finite-dimensional Hilbert space, once an operator is represented as a matrix in a basis, the Hilbert-Schmidt norm is equivalent to the Frobenius norm. It is then clear that the error bound in Eq.~\eqref{eq:matrix_operator_truncation_bound} shares the same form as that of the state in Eq.~\eqref{eq:accuracy_from_rdm_3}, but with respect to the Hilbert-Schmidt norm. As we shall subsequently demonstrate in Section~\ref{subsec:ham_compression}, the NSPO basis offers a compact and efficient way to compress vibrational and vibronic Hamiltonians. It also offers a way to massively reduce the computational effort required to compute the non-zero eigenvalues of the 2-SRDMs, and thus the operator entanglement entropy.

In order to determine which 1-SRDM eigenvectors should be discarded in the construction of the NSPO basis, the Hilbert-Schmidt norm of the operator $\hat{O}$ has to be taken into account. For many-body systems, $||\hat{O}||_{HS}$ can be \textit{extremely} large, and typically increases with the number of degrees of freedom, $d$. As such, the eigenvalue threshold should be set to ensure that $||\bs{O}-\tilde{\bs{O}}||^2_{F} / ||\bs{O}||$ is below a specified accuracy parameter, $\epsilon$, rather that $||\bs{O}-\tilde{\bs{O}}||^2_{F}$ itself. To this end, the eigenvalues $\Lambda_{I}^{(\kappa)}$ of the 1-SRDM $\bs{R}^{(\kappa)}$ are first normalized so that they sum to one. After this, an eigenvalue is taken to be above threshold if $\Lambda_{I}^{(\kappa)} > \epsilon/d$, with $\epsilon$ typically in the range of $10^{-6}$ to $10^{-10}$, depending on the desired level of accuracy.

We end this section by noting connections to other works. First and foremost, it is important to point out that the transformation to a truncated basis of NSPOs is fundamentally the same as the POTFIT algorithm\cite{Jackle1995,Jackle1996} when applied to functions of potential-like functions operators using a discrete variable representation (DVR). In this case, the non-zero part of the 1-SRDMs $\bs{R}^{(k)}$ reduce to the potential density matrices of the POTFIT method. We note, however, that the framework presented here represents a generalization that is applicable to any Hilbert-Schmidt operator, not just potential-type ones. Similarly, the NSPO formalism is also found to fit within the HOSVD framework\cite{doi:10.1137/S0895479896305696, T-HOSVD_2012}, which can be seen to be an later, independent rediscovery of the POTFIT algorithm. We also note that recently the transformation to an operator basis comprising the 1-SRDM eigenvectors was also reported by 
Debertolis\cite{debertolis2025} in the context of exploring the compressibility of the time-evolution and number operators for fermionic Hamiltonians, with application to the study the non-stabilizerness and non-Gaussianity of operators.

\subsection{Super Mutual Information Matrix}

One of the main tools utilized in this work is the extension of the idea of the mutual information (MI) for states $\ket{\Psi} \in \mathcal{H}$ to that of operators $\hat{O} \in B_2(\mathcal{H})$. As mentioned in subsection~\ref{subsec:mi}, the MI matrix maps out bi-partite entanglement across Hilbert subspaces. The only necessary condition on the applicability of the MI is the sub-additivity property of the vNEE, which is a consequence of Klein's inequality theorem\cite{Carlen2009TRACEIA, BHATIA2003125}. The necessary conditions to guarantee that the sub-additivity property holds, and thus the conditions to extend the calculation of MI to an operator $\hat{O}$, are two-fold: (i) that the corresponding SRDMs are Hermitian, and; (ii) the operator admits a spectral decomposition. Full details of the origins of these conditions are given in the Supporting Information. Condition (i) is satisfied by noting that any Gram matrix is Hermitian\cite{Halperin_1962}. Condition (ii) is satisfied by any Hilbert-Schmidt operator since they are compact\cite{Sołtan2018}, as given by the relation in Eq.~\eqref{eq:operator_class_relation}. To retain the same interpretation as that of the standard MI, however, the eigenvalues of the SRDMs are normalized to have unit trace, which is done throughout the paper.

We may now define the operator analogue of the MI, the ``super" MI (SMI), given by

\begin{equation}\label{eq:SMI}
    I_{\kappa\lambda}^{op}=\frac{1}{2}(S_\kappa^{op}+S_\lambda^{op}-S_{\kappa\lambda}^{op})(1-\delta_{\kappa\lambda})\geq0\quad
\end{equation}

\noindent where $S^{op}_\kappa$ is the OEE, which shares the same functional form as the standard vNEE:

\begin{equation}
    S^{op}_\kappa=\sum_{I=1}^{N_\kappa^2}\Lambda_{I}^{(\kappa)}\log\Lambda_{I}^{(\kappa)},
\end{equation}

\noindent
where the $\Lambda_{I}^{(\kappa)}$ are the eigenvalues of the 1-SRDM $\bs{R}^{(\kappa)}$ for degree of freedom $\kappa$. $S^{op}_{\kappa\lambda}$ is constructed entirely analogously, but instead using the eigenvalues of the 2-SRDM $\bs{R}^{(\kappa,\lambda)}$. A subtlety to note is that the OEE $S^{op}_\kappa$ measures the entanglement of operator subspaces, not directly the ability for the operator to generate entanglement on a given state.

By utilizing the operator Hilbert space and constructing the SMI of a given operator $\hat{O}$, we can explicitly examine the entanglement between operator subspaces. At a high level, we interpret the SMI as an operator's ability to act on subspace $A$ without necessarily acting on the subspace $B$. Alternatively, if an operator acts on subspace $A$, the SMI gives a measure of how much the operator also necessarily acts on $B$. This is discussed in more detail in Section~\ref{subsec:smi}, where we analyze the SMI matrices of different vibrational and vibronic Hamiltonians. In the context of quantum dynamics, however, the time-independent Hamiltonian is often not the main operator of interest, but rather the time evolution operator $\hat{U}(\Delta t)=e^{i\hat{H}\Delta t}$. In realistic simulations, it is not usually possible to construct $\hat{U}(\Delta t)$. However, we can obtain a simple approximation of the time evolution operator by considering a first order expansion

\begin{equation}
    \hat{U}^{(1)}(\Delta t):= 1-i\Delta t\hat{H},
\end{equation}

\noindent
which will be valid for short timesteps $\Delta t$. In Section~\ref{subsec:smi} the SMI matrices of $\hat{U}^{(1)}(\Delta t)$ will be discussed for various representative vibronic Hamiltonians.

\section{Calculation of Super Reduced Density Matrices}

We show explicitly how the 1-SRDMs and 2-SRDMs can be efficiently constructed for some common operator representations; namely, for operators represented in the sum-of-product (SOP) and matrix product operator (MPO) formats. These are chosen as they are the most commonly encountered operator formats in molecular quantum dynamics simulations, with SOP format being using in the majority of MCTDH calculations, and MPO format in most MPS simulations.

The 1-SRDMs are typically relatively cheap to compute. The 2-SRDMs, however, can be extremely expensive to compute and store. Moreover, due to their $N^{4} \times N^{4}$ dimension, the calculation of their eigenvalues is prohibitively expensive even for small basis dimensions. As such, the calculation of SMI matrices is inhibited. This may be overcome by first transforming the representation of the operator $\hat{O}$ to the truncated one-mode NSPO bases $\{\elsket{\Upsilon}{\kappa}\}$, which are typically one-to-two orders of magnitude smaller than the full one-mode operator bases $\{ \elsket{I}{\kappa} \}$. The details of this transformation for both SOP and MPO formats is also given.

\subsection{Sum of Product Operators}\label{subsec:nspo_sop_construction}

The SOP format corresponds to the representation of a operator $\hat{O}$ by a sum of tensor products of 1-mode operators:

\begin{equation}
    \hat{O} = \sum_{a=1}^S\mu_a\hat{\nu}_a^{(1)}\otimes\cdots\otimes\hat{\nu}_a^{(d)},
\end{equation}

\noindent where $\mu_{a}\in \mathbb{C}$, and $\hat{\nu}_a^{(\kappa)}\in B_{2}(\mathcal{H^{(\kappa)}})$ is an operator that acts only on the $\kappa^{\text{th}}$ degree of freedom. 

The 1-SRDMs and 2-SRDMs can then be obtained by substituting the SOP form of the operator into Eq.~\eqref{eq:1srdm} and Eq.~\eqref{eq:2srdm}, respectively,



\begin{equation}\label{eq:1srdm_sop}
    \left(R^{(\kappa)}\right)_{ij,\,i'j'}
    = \sum_{a,b=1}^{S} \mu_a\, \mu_b^{*}
    \left( \prod_{\substack{\lambda=1\\ \lambda \neq \kappa}}^{d} \Omega^{(\lambda)}_{ba} \right)
    \left(\nu^{(\kappa)}_a\right)_{ij}
    \left(\nu^{(\kappa)}_b\right)^{\!*}_{i'j'},
\end{equation}

\begin{equation}\label{eq:2srdm_sop}
    \left(R^{(\kappa,\lambda)}\right)_{ijkl,\,i'j'k'l'}
    = \sum_{a,b=1}^{S} \mu_a\, \mu_b^{*}
    \left( \prod_{\substack{\gamma=1\\ \gamma \neq \kappa,\lambda}}^{d} \Omega^{(\gamma)}_{ba} \right)
    \left(\nu^{(\kappa)}_a\right)_{ij}
    \left(\nu^{(\lambda)}_a\right)_{kl}
    \left(\nu^{(\kappa)}_b\right)^{\!*}_{i'j'}
    \left(\nu^{(\lambda)}_b\right)^{\!*}_{k'l'}
\end{equation}

\noindent
where $\bs{\nu}_{a}^{(\kappa)}$ is the matrix representation of the operator $\hat{\nu}_a^{(\kappa)}$ in the basis $\{\ket{i_{\kappa}}\}$ with elements

\begin{equation}
    \left(\nu_a^{(\kappa)}\right)_{ij} = \braket{i_{\kappa} | \hat{\nu}_{a}^{(\kappa)} | j_{\kappa}},
\end{equation}

\noindent
and $\Omega_{ab}^{(\kappa)}$ denotes the Hilbert-Schmidt inner products between operators $\hat{\nu}_{a}^{(\kappa)}$,

\begin{equation}
    \begin{aligned}
        \Omega_{ab}^{(\kappa)} &= \langle \hat{\nu}_{a}^{(\kappa)}|\hat{\nu}_{b}^{(\kappa)}\rangle \\
        & = \tr{\bigl( \hat{\nu}_{a}^{(\kappa)\dagger} \hat{\nu}_{b}^{(\kappa)} \bigr)} \\
        & = \sum_{i,j=1}^{N_{\kappa}} \bigl(\nu_{a}^{(\kappa)}\bigr)_{ij}^{*}
        \bigl({\nu}_{b}^{(\kappa)}\bigr)_{ij}.
    \end{aligned}
\end{equation}

The $\boldsymbol{\Omega}^{(\kappa)}$ are intermediate tensors which are precomputed and stored, making the construction of $\boldsymbol{R}^{(\kappa)}$ over all $\kappa$ more efficient, and are also re-used when constructing the 2-SRDM. The implementation details for constructing the SRDMs and how the intermediate tensors are stored is outlined in the SI, including how the use of a DVR basis, in which many of the matrices $\bs{\nu}_{a}^{(\kappa)}$ are diagonal, can simplify calculations. Writing

\begin{equation}
    \left(\nu_{a}^{(\kappa)}\right)_{ij} = \left(\nu_{a}^{(\kappa)}\right)_{I} = \braket{\hat{I}_{\kappa}|\hat{\nu}_{a}^{(\kappa)}},
\end{equation}

\noindent
and

\begin{equation}
    \begin{aligned}
        \left(\tilde{\nu}^{(\kappa)}_a\right)_{\Upsilon_\kappa}
        &= \langle \hat{\Upsilon}_\kappa \,|\, \hat{\nu}^{(\kappa)}_a \rangle \\
        &= \sum_{I_\kappa=1}^{N_\kappa^{2}} U^{(\kappa)\dagger}_{I_\kappa\Upsilon_\kappa}
        \left(\nu^{(\kappa)}_a\right)_{I_\kappa}.
    \end{aligned}
\end{equation}

\noindent
the representation of $\hat{O}$ in the truncated NSPO basis reads

\begin{equation}
    \label{eq:natsop}
    \tilde{O}_{\Upsilon_{1},\dots,\Upsilon_{d}} \approx \sum_{a=1}^{S} \mu_{a} \left(\tilde{\nu}_{a}^{(1)}\right)_{\Upsilon_{1}} \cdots \left(\tilde{\nu}_{a}^{(d)}\right)_{\Upsilon_{d}},
\end{equation}

\noindent
from which an efficient computation of the 2-SRDMs can be made, as detailed in the Supporting Information. We say that an operator represented in the form of Eq.~\ref{eq:natsop} is in natural SOP format. Importantly, by transforming from SOP format to natural SOP format, the dimension of the $\kappa^{\text{th}}$ 2-SRDM is reduced to $\mathbb{C}^{M_{\kappa}^2\times M_{\kappa}^2}$, significantly reducing the computation cost. To put this into perspective, we observe that a typical molecular quantum dynamics simulation can easily require one-mode vibrational basis sizes of $N_{\kappa}=20$ or more. As we shall detail, typical NSPO basis dimensions for vibrational and vibronic Hamiltonians can be expected to be of the order of 10, and possibly even smaller in many cases. Taking representative values of $N_{\kappa}=20$ and $M_{\kappa}=10$, the dimension of the 2-SRDM matrix would be reduced from 160,000 to 100. As we shall demonstrate via explicit computation in Section~\ref{subsec:ham_compression}, such NSPO basis dimensions are readily achievable in realistic calculations.

\subsection{Matrix Product Operators}\label{subsec:nspo_mpo_construction}

Another widely used operator form is the matrix product operator (MPO) representation. The MPO representation is essential to efficiently compute mappings and expectation values when quantum states are approximated by MPSs. A full review of MPOs is not presented here, and for more details we refer the reader to Refs.~\citenum{SCHOLLWOCK201196,MA202219}. An MPO with open-boundary condition approximates the full operator coefficient tensor as a contraction over a set of rank-four core tensors $\{\bs{W}^{(I_n)}\}$,

\begin{equation}\label{eq:mpos}
\begin{aligned}
    \hat{O} &= \sum_{\alpha_{1}=1}^{w_{1}} \sum_{\alpha_{2}=1}^{w_{2}} \cdots \sum_{\alpha_{d-1}=1}^{w_{d-1}}
    \sum_{i_{1},j_{1}=1}^{N_{1}} \sum_{i_{2},j_{2}=1}^{N_{2}} \cdots \sum_{i_{d},j_{d}=1}^{N_{d}}
    W_{\alpha_{0},\alpha_{1}}^{(i_1,j_1)} W_{\alpha_{1},\alpha_{2}}^{(i_2,j_2)} \cdots W_{\alpha_{d-1},\alpha_{d}}^{(i_d,j_d)} \proj{i_1}{j_1}\otimes\cdots\otimes\proj{i_d}{j_d} \\
    &= \sum_{\alpha_{1}=1}^{w_{1}} \sum_{\alpha_{2}=1}^{w_{2}} \cdots \sum_{\alpha_{d-1}=1}^{w_{d-1}}
    \sum_{I_{1}=1}^{N_{1}^{2}} \sum_{I_{2}=1}^{N_{2}^{2}} \cdots \sum_{I_{d}=1}^{N_{d}^{2}}
    W_{\alpha_{0},\alpha_{1}}^{(I_1)} W_{\alpha_{1},\alpha_{2}}^{(I_2)} \cdots W_{\alpha_{d-1},\alpha_{d}}^{(I_d)}
    \elsket{I}{1}\otimes\cdots\otimes\elsket{I}{d}.
\end{aligned}
\end{equation}

The dimensions of the virtual bonds, $w_i=\dim(\alpha_i)$, are also referred to as bond dimensions, with $w_0=w_{d}=1$ for open-boundary conditions. Identities can be inserted between neighbouring MPO core tensors, which corresponds to a set of gauge freedoms. This freedom enables us to represent the MPO in different canonical forms\cite{SCHOLLWOCK201196,Catarina2023}. This can be exploited to make all but one MPO core orthogonal, the single non-orthogonal MPO core being termed the orthogonality centre. In particular, it is possible to make all MPO cores to the left of the orthogonality center left orthogonal, and those to the right orthogonal. Left orthogonal MPO cores satisfy

\begin{equation}\label{eq:core_tensor_left_orthogonal}
\sum_{\alpha_{\kappa-1},\,i_\kappa,\,j_\kappa}
   \left( W^{(i_\kappa, j_\kappa)}_{\alpha_{\kappa-1},\,\alpha_\kappa} \right)^{\!*}
   W^{(i_\kappa, j_\kappa)}_{\alpha_{\kappa-1},\,\alpha'_\kappa}
= \delta_{\alpha_\kappa,\,\alpha'_\kappa}.
\end{equation}

\noindent
while right orthogonal cores satisfy

\begin{equation}\label{eq:core_tensor_right_orthogonal}
\sum_{\alpha_\kappa,\,i_\kappa,\,j_\kappa}
   W^{(i_\kappa, j_\kappa)}_{\alpha_{\kappa-1},\,\alpha_\kappa}
   \left( W^{(i_\kappa, j_\kappa)}_{\alpha'_{\kappa-1},\,\alpha_\kappa} \right)^{\!*}
= \delta_{\alpha_{\kappa-1},\,\alpha'_{\kappa-1}}.
\end{equation}

\noindent
An MPO with the orthogonality centre placed at position $\kappa$ is referred to as being in mixed orthogonal form. Putting an MPO into mixed orthogonal form, and movements of the orthogonality centre throughout the MPO, can be achieved in practice \textit{via} a series of QR and LU decompositions\cite{McCulloch_2007,SCHOLLWOCK201196}.

Once an MPO is in mixed orthogonal form with the orthogonality centre at position $\kappa$, the 1-SRDM $\bs{R}^{(\kappa)}$ can be easily computed as

\begin{equation}
    R^{(\kappa)}_{ij,\,i'j'}
    = \sum_{\alpha_{\kappa-1}=1}^{w_{\kappa-1}} \sum_{\alpha_\kappa=1}^{w_\kappa}
    W^{(i,j)}_{\alpha_{\kappa-1},\,\alpha_\kappa}
    \left( W^{(i',j')}_{\alpha_{\kappa-1},\,\alpha_\kappa} \right)^{\!*}.
\end{equation}

\begin{figure}
    \centering
    \includegraphics[width=1.0\linewidth]{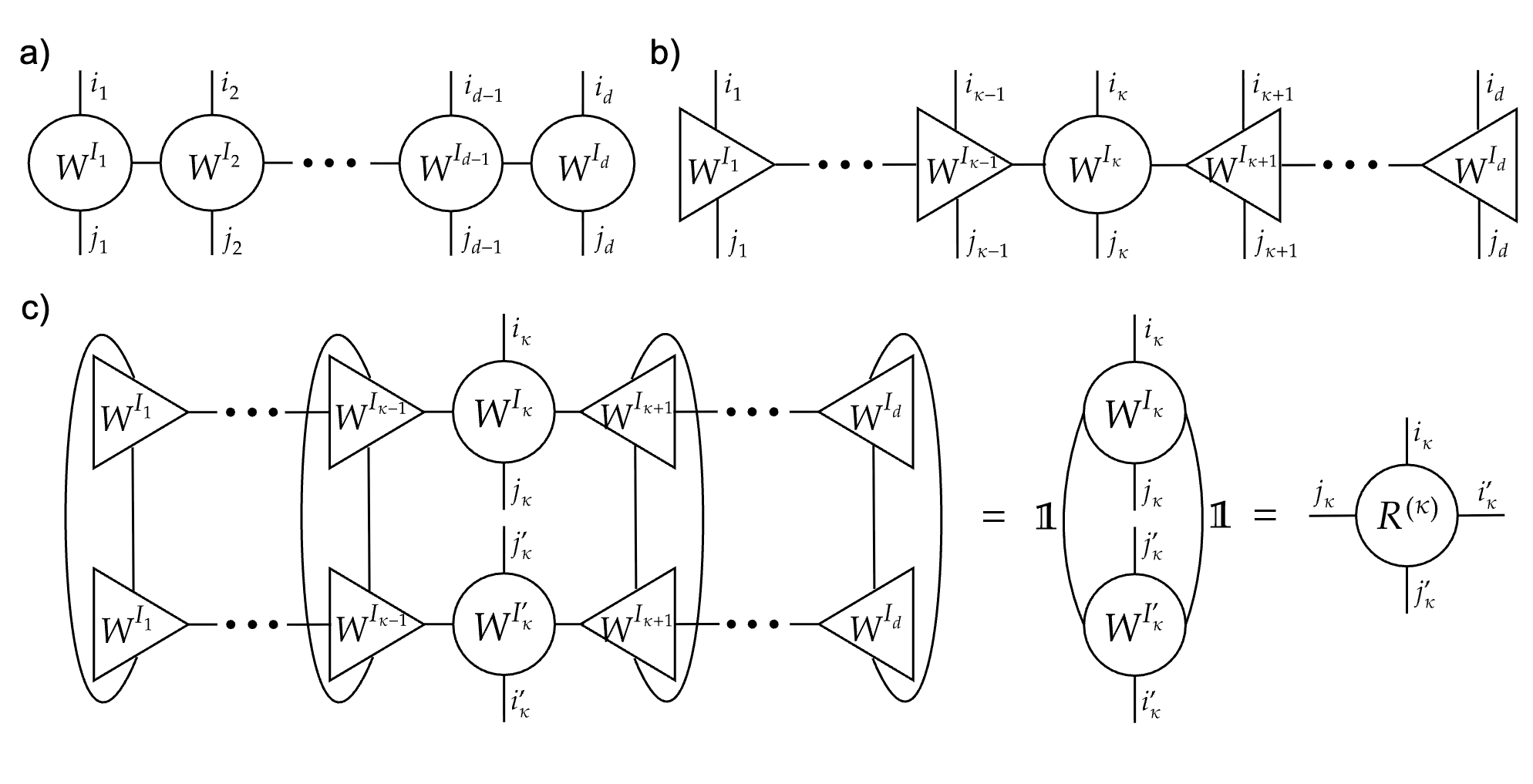}
    \caption{Diagrammatic representation for obtaining the 1-SRDM from MPOs. a) General MPO diagrammatic representation. b) Mixed canonical form with the right (left) pointing triangles representing a left (right) orthogonal tensor. c) Obtaining the 1-SRDM from a mixed canonical form. In this case the traces over the right and left orthogonal tensors evaluates to the identity. All the 1-SRDMs can be obtained by shifting the orthogonality centre down the chain and repeating procedure.}
    \label{fig:mpo_diagram}
\end{figure}

\noindent
The derivation of this is shown diagrammatically Fig.~\ref{fig:mpo_diagram}. All $d$ 1-SRDMs can then be obtained in a single sweep, starting with the orthogonality centre at $\kappa=1$, computing $\bs{R}^{(1)}$, shifting the orthogonality centre to $\kappa=2$, and repeating the procedure.


We now consider the construction of the 2-SRDMs $\bs{R}^{(\kappa,\lambda)}$, where without loss of generality we assume $\kappa < \lambda$. The procedure starts with the orthogonality centre being placed at position $\kappa$. In the evaluation of $\mathrm{Tr}_{B} \elsket{O}{} \elsbra{O}{}$, everything to the left and right of the $\kappa$ and $\lambda$ MPO cores then evaluates to the identity. The 2-SRDM then reduces to

\begin{equation}\label{eq:2srdm_MPO}
    \begin{aligned}
        R^{(\kappa,\lambda)}_{IJ,\,I'J'}
        =\; & \sum_{\alpha_{\kappa-1}=1}^{w_{\kappa-1}}
        \sum_{\alpha_\kappa,\,\alpha'_\kappa=1}^{w_\kappa}
        \sum_{\alpha_{\lambda-1},\,\alpha'_{\lambda-1}=1}^{w_{\lambda-1}}
        \sum_{\alpha_\lambda=1}^{w_\lambda} \\
        & \times W^{(I)}_{\alpha_{\kappa-1},\,\alpha_\kappa}
        \left( W^{(I')}_{\alpha_{\kappa-1},\,\alpha'_\kappa} \right)^{\!*}
        \,\Theta^{(\kappa,\lambda)}_{\alpha_\kappa,\,\alpha'_\kappa,\,\alpha_{\lambda-1},\,\alpha'_{\lambda-1}}\,
        W^{(J)}_{\alpha_{\lambda-1},\,\alpha_\lambda}
        \left( W^{(J')}_{\alpha'_{\lambda-1},\,\alpha_\lambda} \right)^{\!*}.
    \end{aligned}
\end{equation}

\noindent
where

\begin{equation}
    \begin{aligned}
        \Theta^{(\kappa,\lambda)}_{\alpha_\kappa,\,\alpha'_\kappa,\,\alpha_{\lambda-1},\,\alpha'_{\lambda-1}}
        =\; & \sum_{I_{\kappa+1}=1}^{N^{2}_{\kappa+1}} \cdots \sum_{I_{\lambda-1}=1}^{N^{2}_{\lambda-1}}
        \sum_{\alpha_{\kappa+1},\,\alpha'_{\kappa+1}=1}^{w_{\kappa+1}}
        \cdots
        \sum_{\alpha_{\lambda-2},\,\alpha'_{\lambda-2}=1}^{w_{\lambda-2}} \\
        & \times W^{(I_{\kappa+1})}_{\alpha_\kappa,\,\alpha_{\kappa+1}}
        \left(W^{(I_{\kappa+1})}_{\alpha'_\kappa,\,\alpha'_{\kappa+1}}\right)^{\!*}
        \cdots
        W^{(I_{\lambda-1})}_{\alpha_{\lambda-2},\,\alpha_{\lambda-1}}
        \left(W^{(I_{\lambda-1})}_{\alpha'_{\lambda-2},\,\alpha'_{\lambda-1}}\right)^{\!*}.
    \end{aligned}
\end{equation}

\noindent
The intermediate tensor $\Theta^{(\kappa,\lambda+1)}$ can be built from $\Theta^{(\kappa,\lambda)}$ via 

\begin{equation}\label{eq:Theta_recursion}
    \Theta^{(\kappa,\lambda+1)}_{\alpha_\kappa,\,\alpha'_\kappa,\,\alpha_\lambda,\,\alpha'_\lambda}
    = \sum_{I_\lambda=1}^{N^{2}_{\lambda}}
    \sum_{\alpha_{\lambda-1},\,\alpha'_{\lambda-1}=1}^{w_{\lambda-1}}
    \Theta^{(\kappa,\lambda)}_{\alpha_\kappa,\,\alpha'_\kappa,\,\alpha_{\lambda-1},\,\alpha'_{\lambda-1}}\,
    W^{(I_\lambda)}_{\alpha_{\lambda-1},\,\alpha_\lambda}
    \left(W^{(I_\lambda)}_{\alpha'_{\lambda-1},\,\alpha'_\lambda}\right)^{\!*}.
\end{equation}

\noindent
Thus, the set set of 2-SRDMs can be computed in a single sweep as follows. Starting with the orthogonality centre set at $\kappa=1$, the complete subset $\{\bs{R}^{(1,\lambda)}|\lambda=2,\dots,d\}$ can be computed in a single sweep, with the intermediate tensors $\Theta^{(\kappa,\lambda)}$ built recursively using Eq.~\ref{eq:Theta_recursion}. After this, the orthogonality centre is moved to $\kappa=2$, and the subset $\{\bs{R}^{(2,\lambda)}|\lambda=3,\dots,d\}$ is built using the same approach. The whole procedure is repeated until the orthogonality centre is moved to $\kappa=d-1$, at the end of which the set $\{\bs{R}^{(\kappa,\lambda)}|\kappa < \lambda\}$ will have been computed.

\begin{figure}
    \centering
    \includegraphics[width=0.5\linewidth]{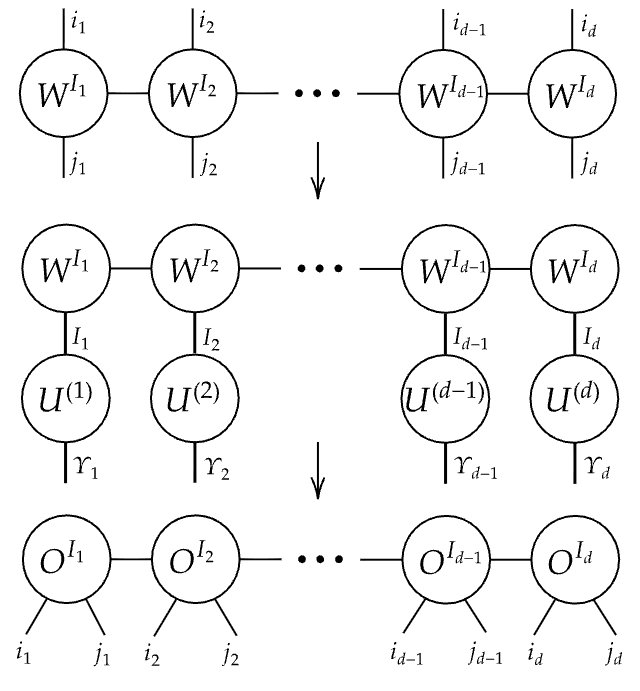}
    \caption{Diagrammatic representation for obtaining the Natural MPO}
    \label{fig:natural_mpo_diagrammatic}
\end{figure}

As with the SOP format case, the calculation of the 2-SRDMs for an operator $\hat{O}$ in MPO format can be massively sped up via first transforming to a truncated NSPO basis. An MPO truncated in the basis of NSPOs, which we call a Natural MPO, can be obtained easily by substituting the MPO core tensor into Eq.~\eqref{eq:1srdm_truncated_coeff_tensor}. This is made simple by considering the diagrammatic representation instead, which is given in Fig.~\ref{fig:natural_mpo_diagrammatic}.

\begin{equation}
    \begin{aligned}
        \tilde{O}_{\Upsilon_1,\ldots,\Upsilon_d}
        &= \sum_{I_1=1}^{N^{2}_{1}} \cdots \sum_{I_d=1}^{N^{2}_{d}}
        O_{I_1,\ldots,I_d}\,
        U^{\dagger}_{I_1\,\Upsilon_1} \cdots U^{\dagger}_{I_d\,\Upsilon_d} \\
        &= \sum_{\boldsymbol{\alpha}}
        \!\left( \sum_{I_1=1}^{N^{2}_{1}} W^{(I_1)}_{\alpha_0,\,\alpha_1}\, U^{\dagger}_{I_1\,\Upsilon_1} \right)
        \cdots
        \!\left( \sum_{I_d=1}^{N^{2}_{d}} W^{(I_d)}_{\alpha_{d-1},\,\alpha_d}\, U^{\dagger}_{I_d\,\Upsilon_d} \right) \\
        &= \sum_{\boldsymbol{\alpha}}
        \tilde{W}^{(\Upsilon_1)}_{\alpha_0,\,\alpha_1} \cdots \tilde{W}^{(\Upsilon_d)}_{\alpha_{d-1},\,\alpha_d}.
    \end{aligned}
\end{equation}

\begin{figure}
    \centering
    \includegraphics[width=1.0\linewidth]{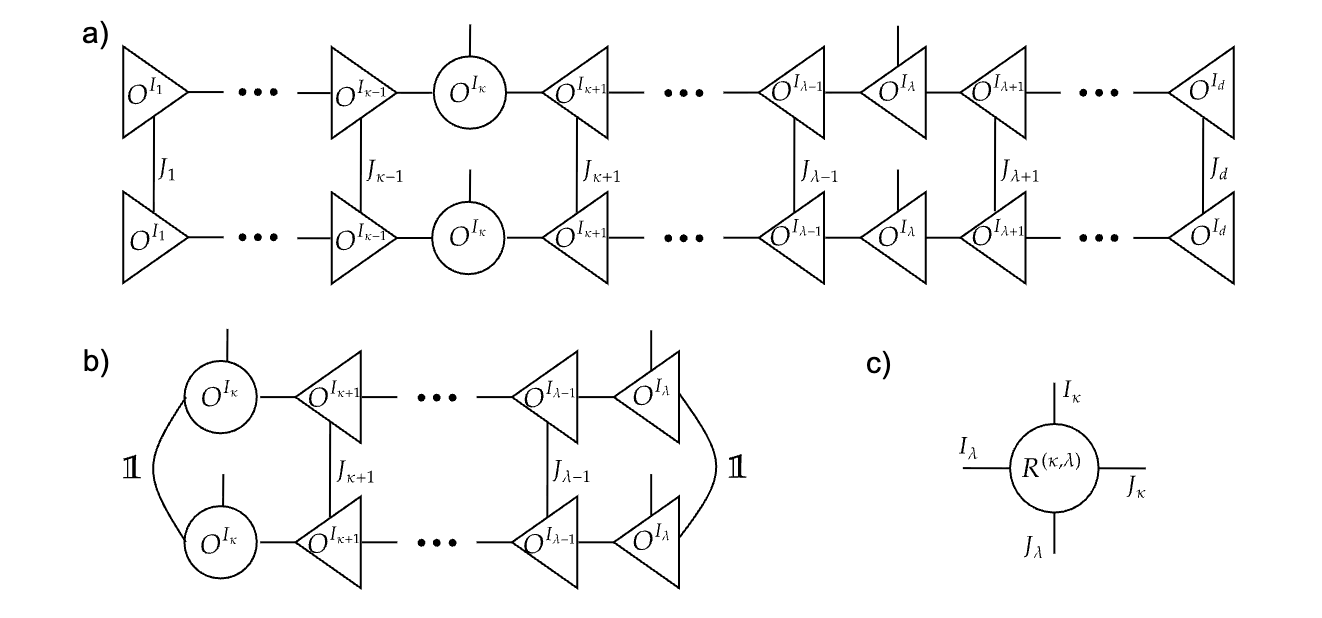}
    \caption{Diagrammatic representation for how the 2-SRDM is constructed from an MPO. a) The orthogonality center is set at one of the degree of freedom. b) The core tensors to the left and right of the $\kappa$ and $\lambda$ degree of freedom evaluates to identity due to orthonormality. The core tensors between $\kappa$ and $\lambda$ needs to be contracted explicitly, but the bond dimension of $J_{\kappa}$ can be significantly less than the primitive operator basis $I_{\kappa}$. c) The 2-SRDM in the NSPO basis has dimensions $R^{(\kappa,\lambda)}\in\mathbb{R}^{M^{\kappa}M^{\lambda}\times M^{\kappa}M^{\lambda}}$}
    \label{fig:mpo_2srdm_diagrammatic}
\end{figure}

\noindent
Using the Natural MPO representation of $\hat{O}$, the 2-SRDMs can be computed by simply replacing the MPO core tensors $\bs{W}$ with the Natural MPO core tensors $\tilde{\bs{W}}$ in Eqs.~\ref{eq:2srdm_MPO} and \ref{eq:Theta_recursion}. This is depicted diagrammatically in Fig.~\ref{fig:mpo_2srdm_diagrammatic}.

Lastly, we note that the MPO format has the additional advantage that the operator Schmidt decomposition (OSD) can be easily obtained. The OSD is defined similarly to the standard Schmidt decomposition, where the operator can be written as a bi-partition between the sub-systems $A$ and $B$ as

\begin{equation}
    \hat{O} = \sum_{i=1}^{m} s_i \hat{A}_i\otimes\hat{B}_i ,
\end{equation}

\noindent where the operator Schmidt basis $\{\hat{A_i}(\hat{B_i})\}$ is orthonormal with respect to the Hilbert-Schmidt inner product, and the total number of non-zero Schmidt coefficients $\{s_i\}$ is called the operator Schmidt number. The exact bond dimension of the $\kappa^{\text{th}}$ MPO core, with the orthogonality center placed at the $\kappa^{\text{th}}$ core, is precisely the Schmidt number of the OSD, where the sub-system $A$ correspond to all the degrees of freedom to the left of the $\kappa^{\text{th}}$ MPO core, and the sub-system $B$ the remaining. Furthermore, the OSD shares the same spectrum as the 1-SRDM, with $s_i^2=\lambda_i$. This relation is particularly useful when analyzing some simple examples of operator compressibility, which is discussed in the following Results section.

\section{Results}

\subsection{Systems Studied}

We consider the calculation of NSPOs and SMI matrices for a number of prototypical vibrational and vibronic model Hamiltonians. Each of these Hamiltonians, comprised of $n_{el}$ electronic states $|\sigma\rangle$ and $n_{v}$ vibrational modes $Q_{\kappa}$, can be written in the following general form

\begin{subequations}
\label{eq:system}
    \begin{align}
        &\hat{H} = \sum_{\sigma,\tau=1}^{n_{el}} \hat{H}_{\sigma,\tau} \otimes |\sigma\rangle\langle\tau|,\label{eq:mol_ham}\\
        &\hat{H}_{\sigma\tau}= \frac{1}{2}\left( \sum_{\kappa=1}^{n_{v}} \omega_{\kappa} \hat{p}_{\kappa}^{2} + E_{\sigma}\hat{1}\right)\delta_{\sigma\tau}
      + \sum_{p=1}^{M}\frac{1}{p!}\sum_{\kappa_{1}=1}^{n_{v}} \sum_{\kappa_{2}=1}^{n_{v}} \cdots \sum_{\kappa_{p}=1}^{n_{v}} \tau_{p,\kappa_{1},\kappa_{2},\dots,\kappa_{p}}^{(\sigma,\tau)} \hat{Q}_{\kappa_{1}} \otimes \hat{Q}_{\kappa_{2}} \otimes \cdots \otimes \hat{Q}_{\kappa_{p}} \label{eq:general_vc_ham},
    \end{align}
\end{subequations}

\noindent
where $\hat{p}_{\kappa}$ and $\hat{Q}_{\kappa}$ are the momentum and position operator for mode $Q_{\kappa}$, respectively, $\omega_{\kappa}$ denote the normal mode frequencies, $E_{\sigma}$ the vertical excitation energies, and the $\tau_{p,\kappa{1},\kappa_{2},\dots,\kappa_{p}}^{(\sigma,\tau)}$ the coupling coefficients.

In this work, we consider a total of four Hamiltonians; two vibronic and two vibrational. The vibronic Hamiltonians correspond to the 24-mode, 2-state pyrazine quadratic vibronic (QVC) Hamiltonian of Raab \textit{et al.}\cite{10.1063/1.478061}, and the 18-mode, 2-state butatriene cation QVC Hamiltonian of Cattarius \textit{et al.}\cite{10.1063/1.1384872}, in which the expansion in Eq.~\ref{eq:general_vc_ham} is truncated at second order ($M=2$). This represents a pair of classic strong vibronic coupling systems containing a conical intersection between excited (ionised) electronic states close to the Franck-Condon point. The two single-state vibrational Hamiltonians correspond to: (i) a sixth-order ground-state vibrational Hamiltonian for ethylene taken from Baiardi \textit{et al.}\cite{doi:10.1021/acs.jctc.7b00329}, and; (ii) a six-dimensional Henon–Heiles Hamiltonian\cite{10.1063/1.3535541}. The ground state ethylene Hamiltonian is derived from Eqs.~\ref{eq:mol_ham} and \ref{eq:general_vc_ham} by setting $n_{el}=1$ and $M=6$. The single-state Henon-Heils Hamiltonian takes the form

\begin{equation}\label{eq:HH_ham}
    \hat{H} = \frac{\omega}{2}\sum_{\kappa=1}^{n_{v}}\left(\hat{p}_{\kappa}^{2} + \hat{Q}_{\kappa}^{2}\right)+\lambda\sum_{\kappa=1}^{n_{v}}\left(\hat{Q}_{\kappa}^{2} \hat{Q}_{\kappa+1}-\frac{1}{3}\hat{Q}_{\kappa}^{3}\right),
\end{equation}

and contains only single-state nearest-neighbor interactions. 

A harmonic oscillator DVR basis is used to represent all the Hamiltonians, with the number of primitive basis for each mode fixed at $N_{k}=21, k={1,\cdots,d}$ for the pyrazine, butatriene and Henon-Heiles Hamiltonian. For the ethylene Hamiltonian the number of primitive is fixed at $N_{k}=11, k={1,\cdots,d}$ following Baiardi \textit{et al.}\cite{doi:10.1021/acs.jctc.7b00329}

\subsection{Hamiltonian Compression}\label{subsec:ham_compression}

\begin{figure}
    \centering
    \includegraphics[width=1.0\linewidth]{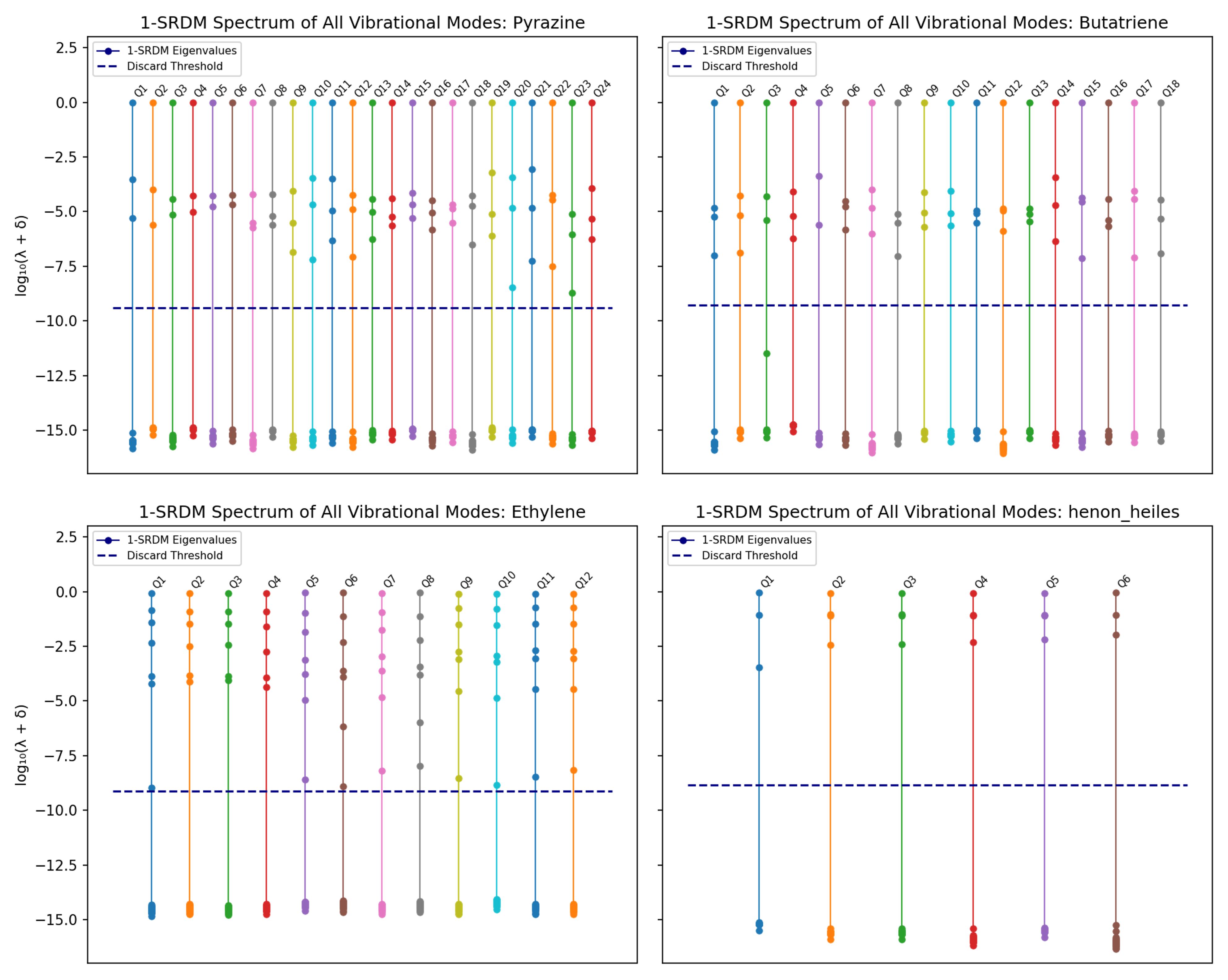}
    \caption{1-SRDM eigenvalue distribution for pyrazine, butatriene, sextic ethylene and Henon-Heiles Hamiltonian. Each vertical line in the plot corresponds to the 1-SRDM for one of the modes, with the mode number labelled at the top. The dashed horizontal line is the threshold value $(\delta=10^{-8}/d)$. The percentage of retained eigenvalues relative to the total number has a maximum $r_{max}$ and minimum $r_{min}$ value of: $r_{max}=0.91\%$ and $r_{min}=0.68\%$ for pyrazine and butatriene, $r_{max}=5.79\%$ and $r_{min}=4.96\%$ for ethylene, and $r_{max}=r_{min}=0.68\%$ for Henon-Heiles}
    \label{fig:1srdm_eigvals}
\end{figure}

As outlined in Section~\ref{sec:op_compression_nspo}, we can truncate a Hamiltonian in the Hilbert-Schmidt norm sense via a transformation of the primitive operator basis $\{\elsket{I}{\kappa}\}$ to the NSPO basis $\{\elsket{\Upsilon}{\kappa}\}$ for each degree of freedom $\kappa$, with only the NSPOs $\elsket{\Upsilon}{\kappa}$ corresponding to significant 1-SRDM eigenvalues retained. To elucidate the compressibility of the types of Hamiltonian typically encountered in molecular quantum dynamics simulations, we show the 1-SRDM eigenvalues for all the vibrational modes for the pyrazine and butatriene cation vibronic coupling Hamiltonians, and the ethylene and Henon-Heiles vibrational Hamiltonians in Fig.~\ref{fig:1srdm_eigvals}.  Each vertical line in Fig.~\ref{fig:1srdm_eigvals} corresponds to the 1-SRDM spectrum for a particular mode, labeled on the top of the spectrum, and the dashed line denotes the cutoff threshold $\delta$ used in the truncation. The truncation threshold for all the following results is chosen to be $\delta=\frac{10^{-8}}{d}$. The spectrum is taken on a log10 scale, shifted by a small constant to avoid problems in regions of the spectrum where the eigenvalues are near zero and can otherwise take negative values due to numerical noise. 

It is evident from Fig.~\ref{fig:1srdm_eigvals} that the 1-SRDM spectrum decays rapidly across all the modes for both the pyrazine and butatriene Hamiltonian, with all the modes requiring no more than four NSPOs. Both of these Hamiltonians have an operator bases $\{\elsket{I}{\kappa}\}$ of dimension 441 per vibrational mode. In contrast, the truncated NSPO bases $\{\elsket{\Upsilon}{\kappa}\}$ do not exceed a dimension of 4. That is, an over 100-fold reduction in the operator basis size, with negligible loss of accuracy in the Hilbert-Schmidt norm sense. We note that the eigenvalues below the truncation threshold are all at least six orders of magnitude lower in value than those above it, and are essentially zero within numerical noise. These results tentatively suggest that vibronic coupling Hamiltonians might in general be extremely highly compressible.

In the Henon-Heiles Hamiltonian, the NSPO basis is exactly 3 dimensional for each degree of freedom, with all remaining 1-SRDM eigenvalues being zero within numerical noise. It is straightforward to understand this result by considering its matrix product operator (MPO) representation. It is known that the MPO representation of a Hamiltonian with nearest-neighbor interaction of the form $\sum_i \hat{X}_i \hat{Y}_{i+1} $ have a maximum bond dimension of three, which can be obtained by using a finite-state automaton\cite{McCulloch_2007,PhysRevA.78.012356,SCHOLLWOCK201196}. For completeness, we also show in the SI explicitly the Henon-Heiles Hamiltonian has an exact constant bond dimension of three using finite-state automaton. As mentioned in Sec.~\ref{subsec:nspo_mpo_construction}, the bond dimension of an MPO between degrees of freedom $\kappa$ and $\kappa+1$ is precisely the Schmidt number of the bi-partitioning of the system into subspaces $1,\dots,\kappa$ and $\kappa+1,\dots,d$. Since the Schmidt coefficients and 1-SRDM eigenvalues are related by $w_i^2=\lambda_i$, the Schmidt number will equal the number of retained 1-SRDM eigenvalues, which for the Henon-Heiles Hamiltonian is upper bounded by three, as seen in Fig.~\ref{fig:1srdm_eigvals}(d).

The ethylene Hamiltonian, although a single-state model, is a high-order sextet expansion of the potential operator, which is significantly higher than the expansion order of two used in the pyrazine and butatriene Hamiltonians. Its 1-SRDM eigenvalue distributions are found to be similar to those of the pyrazine and butatriene vibronic Hamiltonians, with a maximum of only 7 NSPOs retained out of a total of 121. This suggests that the high-level of compressibility seen in the pyrazine and butatriene Hamiltonian is not merely an artifact of their low-order potential expansion order, but is also present in more highly correlated Hamiltonians.

\begin{figure}[]
    \centering
    \includegraphics[width=1.0\linewidth]{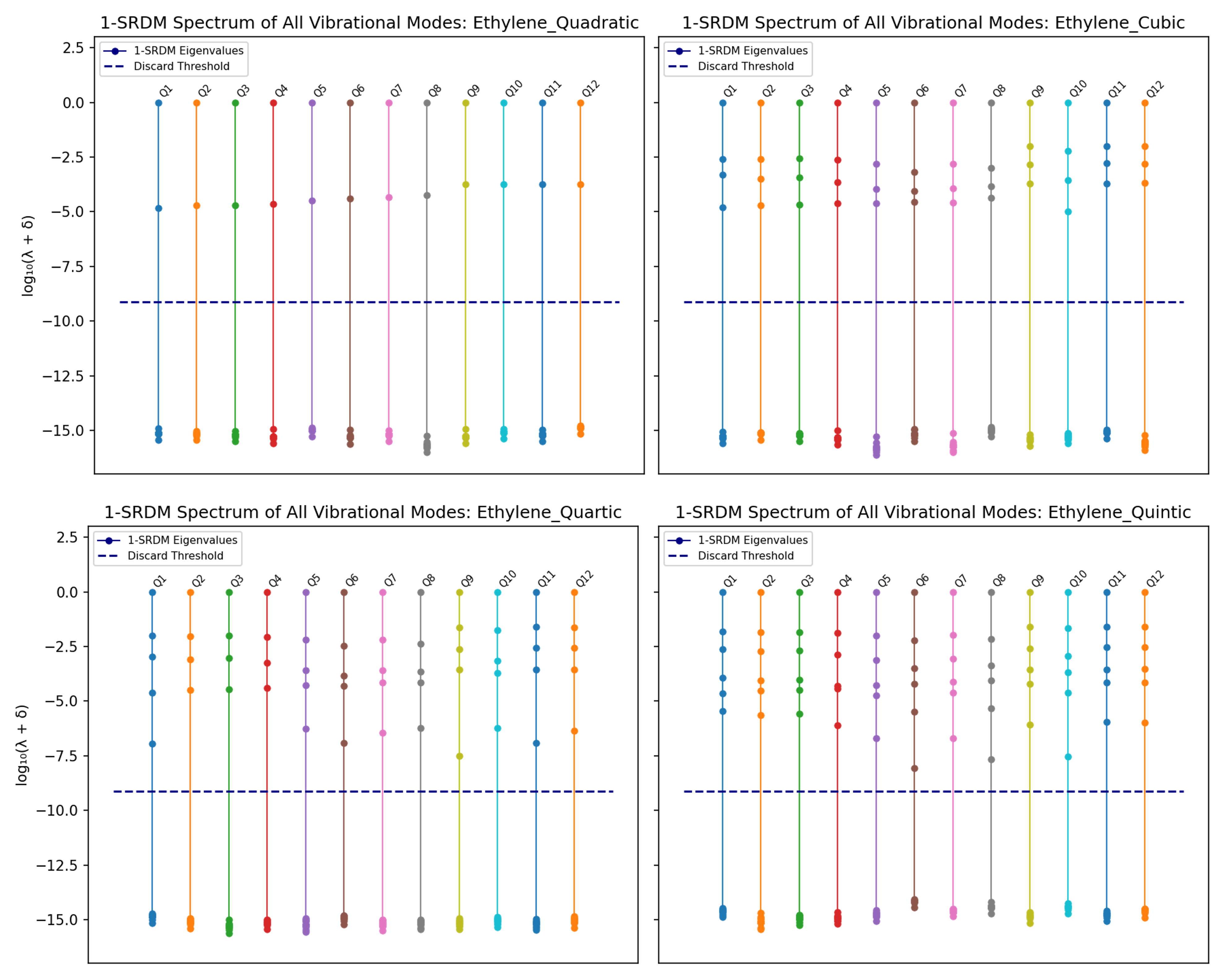}
    \caption{Compressibility of different order ethylene expansion. The percentage of retained eigenvalues relative to the total number has a maximum $r_{max}$ and minimum $r_{min}$ value of: $r_{max}=r_{min}=4.96\%$ for the quintic model, $r_{max}=4.13\%$ and $r_{min}=3.31\%$ for the quartic model, and $r_{max}=r_{min}=3.31\%$ for the cubic model}
    \label{fig:ethylene_order_compressibility}
\end{figure}

We also consider the change in compressibility of the ethylene vibrational Hamiltonian when varying the orders of the potential expansion. Shown in Fig.~\ref{fig:ethylene_order_compressibility} are the 1-SRDM eigenvalue distributions for truncated 2nd, 3rd, 4th, and 5th-order expansions. A general, intuitive trend of higher compressibility (lower number of retained NSPOs) in the lower-order expansions of the Hamiltonian is seen. Fig.~\ref{fig:ethylene_order_compressibility} also anecdotally suggests that, while a lower-order expansion reduces the number of retained NSPOs, the dimension of the truncated NSPO basis does not increase dramatically when using a higher-order expansion. It is worthwhile to note that, by symmetry, there are no terms in the 2nd-order ethylene model coupling the different modes, and this thus correspond to a simple uncoupled Harmonic oscillator Hamiltonian. Using the same finite-state automaton framework and argument as used for the Henon-Heiles Hamiltonian, it can be shown that any Hamiltonian with only single-mode terms, i.e.) $\hat{H}=\sum_{\kappa=1}^d \hat{h}_\kappa$, where $\hat{h}_\kappa \in B_{2}(\mathcal{H})$ only operates on degree of freedom $\kappa$, has a constant bond dimension of 2\cite{McCulloch_2007}. This matches the 2-dimensional NSPO bases for each and every mode in the 2nd-order model.

The compressibility of the Hamiltonian serves an additional, important role in this work, namely in the construction of the 2-SRDMs. The 2-SRDM in the primitive operator basis have dimensions $N_{\kappa}^4\times N_\kappa^4$. Even for the modest $N_{\kappa}=21$ used here, the 2-SRDMs are 194481-dimensional, precluding their calculation and storage in the full operator basis $\{\elsket{I}{\kappa}\}$. In the NSPO basis, however, the largest 2-SRDMs are tiny: 16-dimensional for the pyrazine and butatriene Hamiltonians, and 49-dimensional for the ethylene Hamiltonian. The storage and diagonalisation of these (as is required for the calculating SMI matrices) is thus rendered trivial.


\subsection{Super Mutual Information Matrices}\label{subsec:smi}

\subsubsection{Hamiltonian Super Mutual Information Matrices}
\begin{figure}
    \centering
    \includegraphics[width=1.0\linewidth]{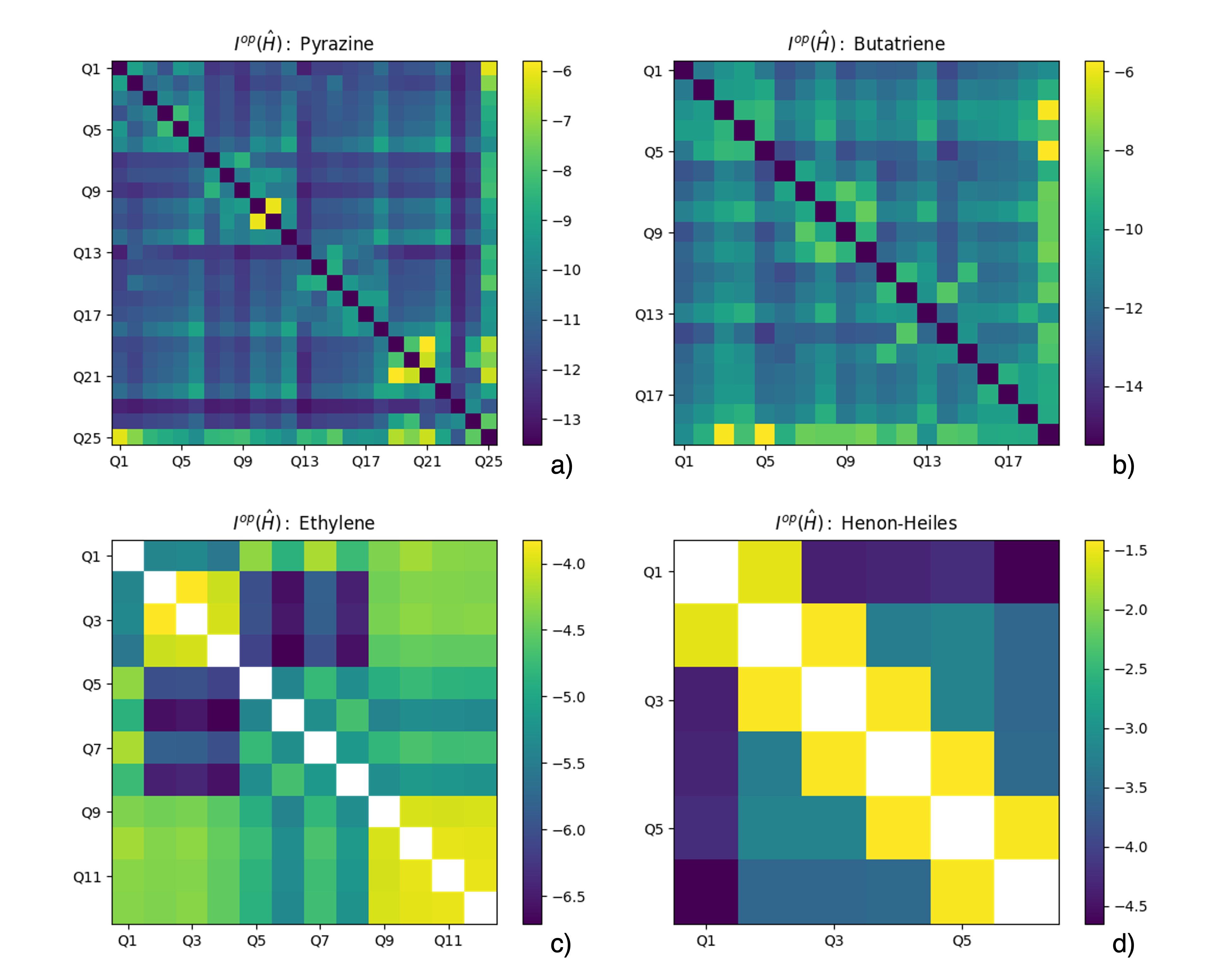}
    \caption{SMI of the Hamiltonian for pyrazine, butatriene, Henon-Heiles, and ethylene. The last degree of freedom in the last row / column corresponds to the electronic degree of freedom for pyrazine and butatriene.}
    \label{fig:log_smis_ham}
\end{figure}

Leveraging the high compressibility of the considered Hamiltonians, we can now efficiently evaluate the 2-SRDMs and their eigenvalues, as required for constructing the SMI. The SMIs using the same set of Hamiltonians from the previous subsection, denoted as $I^{op}(\hat{H})$, are shown in Fig.~\ref{fig:log_smis_ham}, and are also plotted on a log10 scale shifted by a small constant to prevent taking the logarithm of negative values. The $\kappa^{\text{th}}$ row and column of the SMI matrix corresponds to the $\kappa^{\text{th}}$ degree of freedom, with the final row and column corresponding to the electronic degree of freedom in the pyrazine and butatriene examples.

Analogous to the standard MI matrix, which maps out the pair-wise correlation between Hilbert subspaces of a quantum state, the SMI matrix enables us to examine the pair-wise correlation between one-particle operators acting on each degree of freedom. That is, the elements $I_{\kappa\lambda}$ in Fig.~\ref{fig:log_smis_ham} measure the correlation between the local operators acting on the $\kappa^{\text{th}}$ and $\lambda^{\text{th}}$ Hilbert subspaces $\mathcal{H}^{(\kappa)}$ and $\mathcal{H}^{(\lambda)}$ in the Hamiltonian. The SMI of the Henon–Heiles Hamiltonian shown in Fig.~\ref{fig:log_smis_ham}(d) is perhaps the most straightforward to interpret. Its predominantly tri-diagonal structure can be understood intuitively from the nearest-neighbor interaction in the Hamiltonian. However, we also observe a weaker, indirect set of couplings coupling between the local operators acting on non-neighboring modes that decays with the distance between them. This represents information that cannot be directly extracted from the parameters of the model Hamiltonian alone.

The SMI also highlights the degrees of freedom that are most strongly coupled. For example, we can see from the SMI matrix for pyrazine in Fig~\ref{fig:log_smis_ham}(a) that the electronic degree of freedom (corresponding to the last column and row) has the highest SMI value with the $Q_1$ vibrational mode which, by symmetry, is the only first-order interstate (off-diagonal) coupling mode in this system. That is, the single vibrational degree of freedom for which the interstate coupling term $\hat{Q}_{\kappa} \otimes |1\rangle\langle2|$ exists in the Hamiltonian. We also observe that both pairs of modes, $(Q_{10},Q_{11})$ and $(Q_{21},Q_{19})$, have large SMI values, as they correspond to strongly bi-linearly coupled modes pairs. That is, there exists terms in the Hamiltonian of the form $\hat{Q}_{\kappa} \otimes \hat{Q}_{\lambda} \otimes |\sigma\rangle\langle\sigma|$, $\kappa \ne \lambda$, with significant prefactors for these pairs of modes. Similarly, in the butatriene cation Hamiltonian, the electronic degree of freedom has the highest SMI value with the single first-order coupling mode $Q_5$, as shown in Fig~\ref{fig:log_smis_ham}(b). We also see that the electronic degree of freedom exhibits high SMI values with modes $Q_3$ and $Q_4$, reflecting that fact that the interstate bi-linear coupling term $\hat{Q}_{3} \otimes \hat{Q}_{4} \otimes |1\rangle\langle2|$ contributes significantly to the Hamiltonian.

\subsubsection{Approximate Time-Evolution Operator Super Mutual Information Matrices}
As the time-evolution operator is often the main interest in quantum dynamics, we consider next the SMI formed from the first-order expansion of the time evolution operator,

\begin{equation}
\hat{U}(\Delta t) \approx \hat{U}^{(1)}(\Delta t):= \hat{1}-i\Delta t\hat{H}.
\end{equation}

\noindent
Here $\Delta t$ is set to $\Delta t=0.1\text{ a.u.}$ in order to ensure the validity of the first-order approximation. 

The time evolution operator generated by a Hamiltonian of uncoupled single-mode terms, $\hat{H}=\sum_{\kappa=1}^d \hat{h}_\kappa$, has an operator Schmidt number of one along any bi-partition, whereas the Hamiltonian itself has an operator Schmidt number of two. That is, the time evolution operator in this case is expressible as a single tensor product,

\begin{equation}\label{eq:U_of_sum_of_local_ham}
    \begin{aligned}
        \hat{U} &= e^{i\hat{H}t}=e^{i\sum_\kappa\hat{h}_\kappa t} \\
        & = e^{i\hat{h}_1 t}\otimes\cdots\otimes e^{i\hat{h}_dt} \\
        &= \hat{U}_1\otimes\hat{U}_2\otimes\cdots\otimes\hat{U}_d ,
    \end{aligned}
\end{equation}

\noindent
where in the second line of Eq.~\eqref{eq:U_of_sum_of_local_ham} the commutation of local operators $[\hat{h}_\kappa,\hat{h}_\lambda]=0, \kappa \neq \lambda$ is used. Consequently, the Hamiltonian will have non-zero SMI matrix elements, whilst the exact time evolution operator will have a zero matrix for the SMI. However, the truncation of the time evolution operator at first-order introduces artifacts in the operator Schmidt number as $\hat{1}-i\Delta t\hat{H}$ does not have a product form. To this end, we instead consider an \textit{ad hoc} ``corrected" modification of $\hat{U}^{(1)}(\Delta t)$ that generates a zero SMI matrix in the limit of an uncorrelated Hamiltonian $\hat{H}$. Specifically, we consider the difference between the SMI formed from $\hat{U}^{(1)}(\Delta t)$ and the SMI formed from 

\begin{equation}
    \hat{U}^{(1)}_{uncorr}(\Delta t)= 1-i\Delta t\hat{H}_{uncorr},
\end{equation}

\noindent
where $\hat{H}_{uncorr}$ represents the uncorrelated part of the Hamiltonian. That is, the sum of only the terms in Eqs.~\ref{eq:mol_ham} and \ref{eq:general_vc_ham} corresponding to a single electronic state and a single vibrational mode. Let the SMI matrices of $\hat{U}^{(1)}$ and $\hat{U}_{uncorr}^{(1)}$ be denoted by $I^{op}(\hat{U}^{(1)})$ and $I^{op}(\hat{U}_{uncorr}^{(1)})$, respectively. Then, in the following discussion, we shall denote by $I_{corr}^{op}(\hat{U}^{(1)})$ the difference

\begin{equation}
    I_{corr}^{op}(\hat{U}^{(1)}) = I^{op}(\hat{U}^{(1)}) -I^{op}(\hat{U}_{uncorr}^{(1)}),
\end{equation}

\noindent
the subscript ``corr" standing for ``corrected".

\begin{figure}[]
    \centering
    \includegraphics[width=1.0\linewidth]{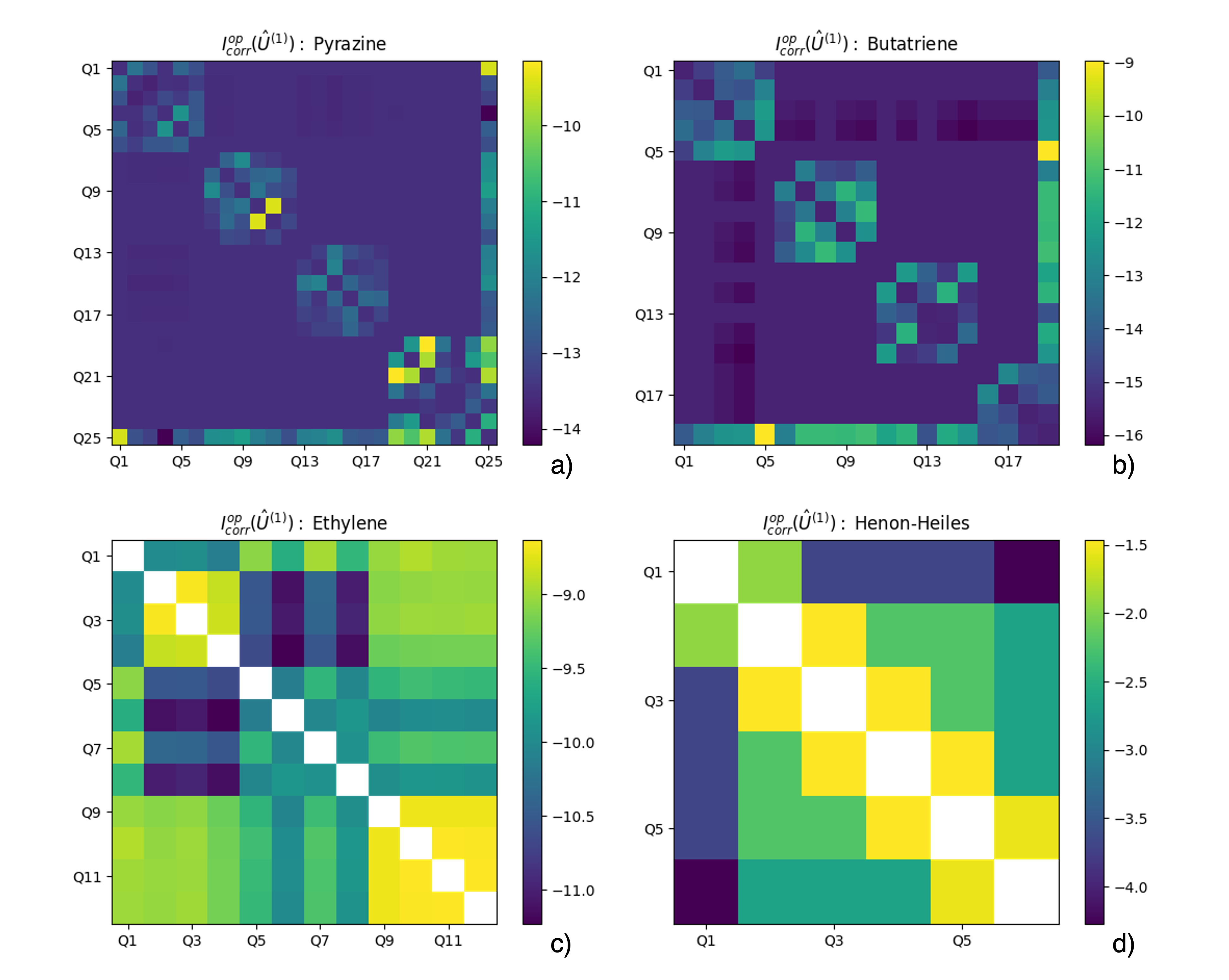}
    \caption{SMI of $\hat{U}^{(1)}_{corr}(\Delta t)$ for pyrazine, butatriene, Henon-Heiles, and ethylene. The last degree of freedom in the last row / column corresponds to the electronic degree of freedom for pyrazine and butatriene.}
    \label{fig:log_smis_Ucorr}
\end{figure}

The corrected SMIs $I_{corr}^{op}(\hat{U}^{(1)})$ for each Hamiltonian is in Fig.~\ref{fig:log_smis_Ucorr}. It is found that the $I_{corr}^{op}(\hat{U}^{(1)})$ show some similarities to the corresponding Hamiltonian SMIs, such as the dominance of the elements corresponding to the electronic degree of freedom and the single first-order coupling modes in each case. However, there are also important differences. In particular, as seen in the Pyrazine (Fig.~\ref{fig:log_smis_Ucorr}(a)) and Butatriene (Fig.~\ref{fig:log_smis_Ucorr}(b)) corrected SMIs, we observe dominant intra-vibrational blocks corresponding to the subsets of mode irreducible representation (irrep) pairs $(\Gamma^{\kappa}, \Gamma^{\lambda})$ such that $\Gamma^{\kappa} \otimes \Gamma^{\lambda} = \Gamma^{|1\rangle} \otimes \Gamma^{|2\rangle}$, where $\Gamma^{|1\rangle}$ and $\Gamma^{|2\rangle}$ are the irreps of the two vibronically coupled electronic states. That is, for pairs of mode symmetries that allow, by symmetry, for bi-linear coupling of the two electronic states via terms of the form $\hat{Q}_{\kappa} \otimes \hat{Q}_{\lambda} \otimes |1\rangle \langle2|$, $\kappa \ne \lambda$.

Both pyrazine and butatriene possess $D_{2h}$ symmetry at the Franck-Condon point. For pyrazine, the two electronic states generate the irreps $\Gamma^{|1\rangle}=B_{3u}$ and $\Gamma^{|2\rangle}=B_{2u}$. The dominant inter-vibrational blocks, from the top left to the bottom right, correspond to the pairs of mode irreps $(A_g, B_{1g})$, $(B_{2g}, B_{3g})$, $(A_u, B_{1u})$ and $(B_{3u}, B_{2u})$, respectively. That is, the pairs of irreps whose direct product yields $B_{3u} \otimes B_{2u} = B_{1g}$ and can bi-linearly couple the two electronic states by symmetry. For butatriene, the two electronic states generate the irreps $\Gamma^{|1\rangle}=B_{2g}$ and $\Gamma^{|2\rangle}=B_{2u}$. The dominant inter-vibrational blocks of the corrected SMI $I_{corr}^{op}(\hat{U}^{(1)})$ in this case correspond the pairs of irreps whose direct products give $B_{2g} \otimes B_{2u}=A_{u}$. That is, they correspond to the mode irrep pairs $(A_u, A_{g})$, $(B_{2g}, B_{2u})$, $(B_{3u}, B_{3g})$ and $(B_{1g}, B_{1u})$. 

\begin{figure}
    \centering
    \includegraphics[width=1.0\linewidth]{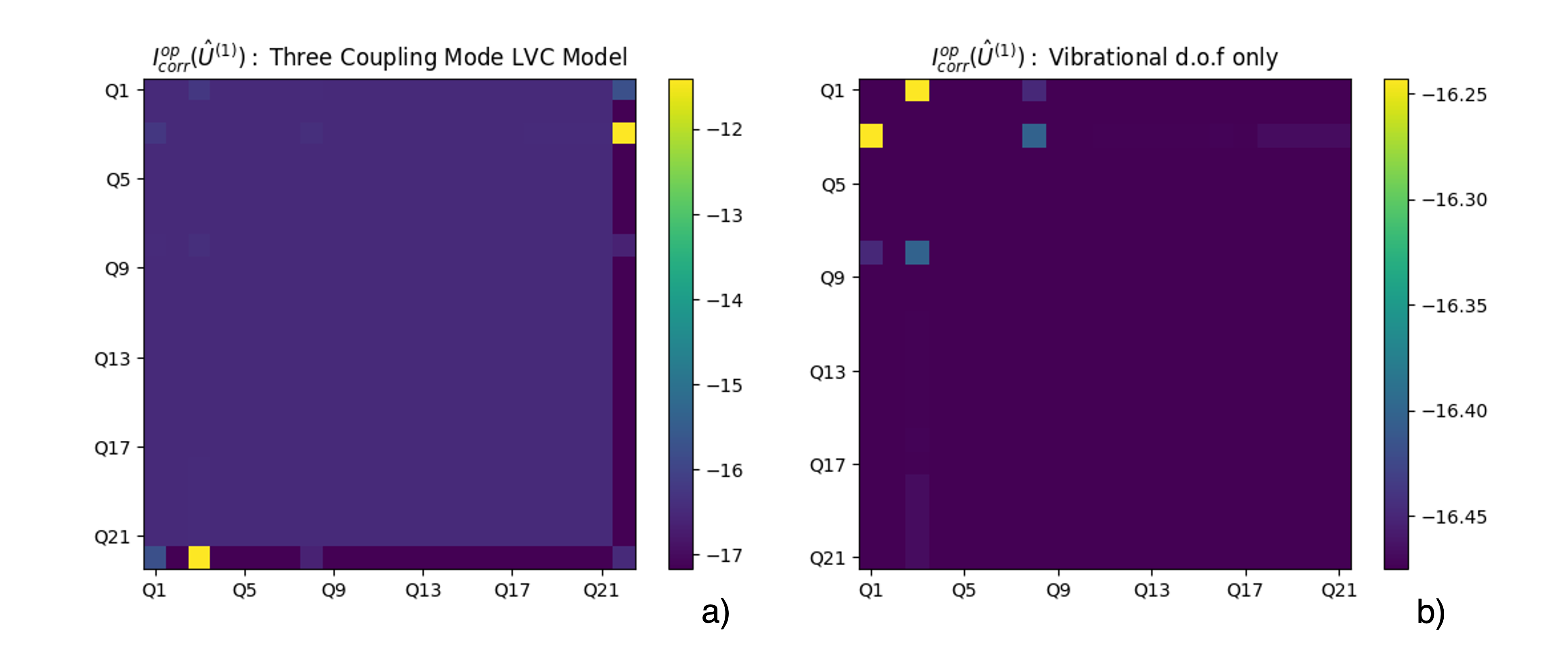}
    \caption{SMI of $U^{(1)}_{corr}$ for the toy LVC model including (a) and excluding (b) the electronic degree of freedom.}
    \label{fig:three_Qc_lvc}
\end{figure}

Another interesting feature we can observe in the corrected SMIs $I_{corr}^{op}(\hat{U}^{(1)})$ are the indirect coupling of modes through their coupling to the electronic degree of freedom. This is made explicit by considering a 21 mode, two-state toy linear vibronic coupling (LVC) Hamiltonian with three coupling modes $Q_1,Q_3$ and $Q_8$ modes, 

\begin{equation}
    \hat{H}_{\sigma\tau}= \frac{1}{2}\left[ \sum_{\kappa=1}^{n_{v}} \omega_{\kappa} \left(\hat{Q}_{\kappa}^{2} + \hat{p}_{\kappa}^{2} \right) + E_{\sigma}\hat{1}\right] \delta_{\sigma\tau}
     + \sum_{\kappa=1}^{n_{v}} \left[ \kappa_{\kappa}^{(\sigma)} \hat{Q}_{\kappa} \delta_{\sigma\tau} + \lambda_{\kappa}\hat{Q}_{\kappa} (1-\delta_{\sigma\tau}) \right],
\end{equation}

\noindent
with inter-state coupling coefficients of $\lambda_{1}=-0.007$~eV, $\lambda_{3}=0.07$~eV and $\lambda_{8}-0.003$~eV. The complete set of parameter and coupling values for the LVC model is included in the SI. Note that in this model, there exists no direct coupling between the vibrational modes $Q_{\kappa}$. However, the corresponding corrected SMIs $I_{corr}^{op}(\hat{U}^{(1)})$, shown in Fig.~\ref{fig:three_Qc_lvc}, clearly shows the indirect coupling of operators acting on the three modes $Q_1,Q_3$ and $Q_8$ via their coupling to the electronic degree of freedom. This is most easily seen in Fig.~\ref{fig:three_Qc_lvc}(b), in which the electronic degree of freedom has been omitted. The magnitude of these SMI elements is significantly smaller than those corresponding to the direct vibrational-electronic coupling, however, as seen in Fig.~\ref{fig:three_Qc_lvc}(a). However, that this indirect coupling is visible is testament to the utility of the SMI approach.

We note that, in the case of the physically-motivated model Hamiltonians considered here, much of the information about the direct coupling of local operators that are encoded in the SMIs can also be extracted from an analysis of the coupling coefficient values. However, such simple models are limited in their scope. A more general, and arguably more powerful, approach is to convert an accurate, but black-box, machine-learned Hamiltonian into a low-rank tensor form that is compatible with quantum dynamics simulations. Examples include the Monte Carlo Canonical Polyadic Decomposition (MCCPD)\cite{Schroder2020}, Monte Carlo Potfit\cite{Schroder2017}, and Neural Network Matrix Product Operator (NN-MPO)\cite{Kentaro2025} methods. Here, the resulting Hamiltonians are in the form of a tensor factorisation and are not amenable to such a simple analysis. In this case, the SMI opens the way to a more universal, formally rigorous and automated way to quantify correlations in operators, including those that are indirect in nature.

It is important to emphasize, however, that the OEE is not equivalent to the entanglement created by the operator, and can be extended to any operator in $B_2(\mathcal{H})$. The SMI is therefore necessary when the entanglement needs to be analyzed at the operator level. While the Hamiltonian and the time-evolution operator SMI analyzed here was motivated by the desire to gauge correlations between degrees of freedom that will manifest in the resulting time-evolved quantum state, this motivation does not extend to general operators that may be useful in dynamical settings. Importantly, the operator Hilbert space encodes additional information that remains relatively unexplored and may contain useful information that is otherwise not directly accessible from state-based analyses alone.

\section{Outlook}

In the present work, we have reviewed the theory of operator entanglement entropy within the framework of operator Hilbert space, which enabled us to rigorously extend quantum information-based entanglement measures, typically associated with quantum states, to operators in finite-dimensional variational methods. Specifically, we introduced the NSPOs, defined as the 1-SRDM eigenvectors, and the SMI, defined with respect to the operator entanglement entropy. Practical approaches to computing these quantities have also been presented for the commonly-used SOP and MPO operator formats. We have shown one application of the NSPOs as a way to compress an operator and enable an extremely efficient operator representation. This is demonstrated for a set of realistic benchmark Hamiltonians used in molecular quantum dynamics, namely a pair of quadratic vibronic coupling Hamiltonians and a pair of single-state vibrational Hamiltonians. For all the Hamiltonians studied in this work, we observe high compressibility and an order of 100 times reduction in the one-mode operator basis dimensions. The SMIs for the same set of Hamiltonians and the associated first-order time-evolution operators is also presented. In these physcially motivated models, we find that the SMIs reveal not just information about direct coupling between local operators that is also present in the corresponding coupling coefficient values, but also information about indirect couplings that cannot otherwise be extracted. Perhaps more importantly, the SMI approach offers a rigorous, quantitative and fully automated way to quantify correlations in operators that are expressed in the general, but opaque, form of a low-rank tensor factorisation, and which may not otherwise be easily interpretable.

\begin{acknowledgement}
M.S.S. thanks the Natural Sciences and Engineering Research Council (NSERC) Alliance and Discovery grant programs as well as NRC Applied Quantum Computing projects (NRC-AQC-111, NRC-AQC-202) for financial support, and all authors thank Aaron Goldberg for helpful discussions.
\end{acknowledgement}



\bibliography{opt}

\end{document}


\renewcommand{\theequation}{S\arabic{equation}}
\renewcommand{\thefigure}{S\arabic{figure}}
\renewcommand{\thetable}{S\arabic{table}}

\section{Calculation of the 1-SRDMs and 2-SRDMs for operators in sum-of-product format}

\subsubsection{General forms}
Let

\begin{equation}
  \hat{O} \;=\; \sum_{a=1}^{S} \mu_a\,
                \hat{\nu}_a^{(1)} \otimes \hat{\nu}_a^{(2)} \otimes \cdots
                \otimes \hat{\nu}_a^{(d)},
  \label{eq:sop}
\end{equation}

\noindent
denote an operator in sum-of-products (SOP) form. The representation of $\hat{O}$ in the basis

\begin{equation}
    \bigotimes_{\kappa=1}^{d} \{ |i_{\kappa}\rangle \langle j_{\kappa} | \}
\end{equation}

\noindent
reads

\begin{equation}
O_{i_1,i_2,\ldots,i_d;\,j_1,j_2,\ldots,j_d}
\;=\; \sum_{a=1}^{S} \mu_a
   \left(\nu_a^{(1)}\right)_{i_1 j_1}
   \left(\nu_a^{(2)}\right)_{i_2 j_2}
   \cdots
   \left(\nu_a^{(d)}\right)_{i_d j_d}.
\end{equation}

\noindent
where

\begin{equation}
    \bigl(\nu_a^{(\kappa)}\bigr)_{i_{\kappa}j_{\kappa}} = \left\langle i_{\kappa} \middle| \hat{\nu}_{a}^{(\kappa)} \middle| j_{\kappa} \right\rangle
\end{equation}

The 1-SRDMs are given by

\begin{equation}
R^{(\kappa)}_{ij;\,i'j'}
\;=\; \sum_{a,b=1}^{S}
   \Gamma^{(\kappa)}_{ab}\,
   \left(\nu_a^{(\kappa)}\right)_{ij}
   \left(\nu_b^{(\kappa)}\right)^{\!*}_{i'j'} .
\end{equation}

\noindent
and the 2-SRDMs by

\begin{equation}
R^{(\kappa,\lambda)}_{ijkl;\,i'j'k'l'}
\;=\; \sum_{a,b=1}^{S}
   \Gamma^{(\kappa,\lambda)}_{ab}\,
   \left(\nu_a^{(\kappa)}\right)_{ij}
   \left(\nu_a^{(\lambda)}\right)_{kl}
   \left(\nu_b^{(\kappa)}\right)^{\!*}_{i'j'}
   \left(\nu_b^{(\lambda)}\right)^{\!*}_{k'l'} .
\end{equation}

\noindent
where

\begin{equation}
\Gamma^{(\kappa)}_{ab}
\;=\; \mu_a\,\mu_b^{*}
   \prod_{\lambda \neq \kappa} \Omega^{(\lambda)}_{ba} .
\end{equation}

\begin{equation}
\Gamma^{(\kappa,\lambda)}_{ab}
\;=\; \mu_a\,\mu_b^{*}
   \prod_{\gamma \neq \kappa,\lambda} \Omega^{(\gamma)}_{ba} .
\end{equation}

\noindent
with

\begin{equation}
\Omega^{(\lambda)}_{ab}
\;=\; \sum_{i,j=1}^{N_\lambda}
   \left(\nu_a^{(\lambda)}\right)^{\!*}_{ij}
   \left(\nu_b^{(\lambda)}\right)_{ij} .
\end{equation}

\subsection{Efficient calculation of the $\Omega$-tensor}
In practice, many of the factor matrices $\boldsymbol{\nu}_{a}^{(\kappa)}$ may be either unit or diagonal matrices. Unit factors, for example, are prevalent in the widely used vibronic coupling Hamiltonian model: if an $n$th-order expansion of the diabatic potential matrix elements is used, then each term $\mu_{a} \hat{\nu}_{a}^{(1)} \otimes \hat{\nu}_{a}^{(2)} \otimes \cdots \otimes \hat{\nu}_{a}^{(d)}$ will contain at a minimum of $d-n$ unit operators $\hat{\nu}_{a}^{(\kappa)} = \hat{1}^{(\kappa)}$. Diagonal factors occur, for example, when representing a potential-like operator $\hat{\nu}_{a}^{(\kappa)}$ in a discrete variable representation (DVR) basis. Taking this into account can lead to substantial computational gains.

Let each factor matrix $\boldsymbol{\nu}_{a}^{(\kappa)}$ be labeled by ``full", ``diagonal", or ``unit", depending on whether it is a dense, digaonal or unit matrix, respectively. Then, the calculation of the a single element $\Omega_{ab}^{(\kappa)}$ can be computed with a minimum floating point operation (FLOP) count according to the the working equations given in Table~\ref{tab:omega}

\begin{table}[h]
\centering
\caption{Closed-form expressions for the matrix element
$\Omega^{(\kappa)}_{ab} \;=\; \sum_{i,j=1}^{N_\kappa}
   \bigl(\nu_a^{(\kappa)}\bigr)^{\!*}_{ij}\bigl(\nu_b^{(\kappa)}\bigr)_{ij}$,
classified by the storage type of each factor.
A ``unit'' factor is the $N_\kappa \times N_\kappa$ identity and is not stored explicitly.
A ``diagonal'' factor is stored as the length-$N_\kappa$ vector $v_a^{(\kappa)}$
of its diagonal entries, $\bigl(v_a^{(\kappa)}\bigr)_i \equiv \bigl(\nu_a^{(\kappa)}\bigr)_{ii}$.
A ``full'' factor is stored as the dense matrix
$\nu_a^{(\kappa)} \in \mathbb{C}^{N_\kappa \times N_\kappa}$.
Only the (full, full) case incurs $\mathcal{O}(N_\kappa^{2})$ work;
every other branch is at most linear in $N_\kappa$.}
\label{tab:omega}
\begin{tabular}{lll}
\toprule
($a$-type, $b$-type) & $\Omega^{(\kappa)}_{ab}$ & FLOP scaling \\
\midrule
(unit, unit)
  & $N_\kappa$
  & $\mathcal{O}(1)$ \\[3pt]
(unit, diagonal)
  & $\displaystyle \sum_{i=1}^{N_\kappa} \bigl(v_b^{(\kappa)}\bigr)_{i}$
  & $\mathcal{O}(N_\kappa)$ \\[8pt]
(unit, full)
  & $\displaystyle \sum_{i=1}^{N_\kappa} \bigl(\nu_b^{(\kappa)}\bigr)_{ii}$
  & $\mathcal{O}(N_\kappa)$ \\[8pt]
(diagonal, unit)
  & $\displaystyle \sum_{i=1}^{N_\kappa} \bigl(v_a^{(\kappa)}\bigr)^{\!*}_{i}$
  & $\mathcal{O}(N_\kappa)$ \\[8pt]
(diagonal, diagonal)
  & $\displaystyle \sum_{i=1}^{N_\kappa} \bigl(v_a^{(\kappa)}\bigr)^{\!*}_{i}\bigl(v_b^{(\kappa)}\bigr)_{i}$
  & $\mathcal{O}(N_\kappa)$ \\[8pt]
(diagonal, full)
  & $\displaystyle \sum_{i=1}^{N_\kappa} \bigl(v_a^{(\kappa)}\bigr)^{\!*}_{i}\bigl(\nu_b^{(\kappa)}\bigr)_{ii}$
  & $\mathcal{O}(N_\kappa)$ \\[8pt]
(full, unit)
  & $\displaystyle \sum_{i=1}^{N_\kappa} \bigl(\nu_a^{(\kappa)}\bigr)^{\!*}_{ii}$
  & $\mathcal{O}(N_\kappa)$ \\[8pt]
(full, diagonal)
  & $\displaystyle \sum_{i=1}^{N_\kappa} \bigl(\nu_a^{(\kappa)}\bigr)^{\!*}_{ii}\bigl(v_b^{(\kappa)}\bigr)_{i}$
  & $\mathcal{O}(N_\kappa)$ \\[8pt]
(full, full)
  & $\displaystyle \sum_{i,j=1}^{N_\kappa} \bigl(\nu_a^{(\kappa)}\bigr)^{\!*}_{ij}\bigl(\nu_b^{(\kappa)}\bigr)_{ij}$
  & $\mathcal{O}(N_\kappa^{2})$ \\
\bottomrule
\end{tabular}
\label{tab:omega}
\end{table}

\begin{equation}
\begin{aligned}
\mathrm{Tr}\bigl(\rho_{AB}\log(\rho_A\otimes\rho_B)\bigr)
&= \mathrm{Tr}\Bigl(\rho_{AB}\,\bigl[\log(\rho_A)\otimes\mathbb{1} + \mathbb{1}\otimes\log(\rho_B)\bigr]\Bigr) \\
&= \mathrm{Tr}\bigl(\rho_{AB}\,[\log(\rho_A)\otimes\mathbb{1}]\bigr)
   + \mathrm{Tr}\bigl(\rho_{AB}\,[\mathbb{1}\otimes\log(\rho_B)]\bigr) \\
&= \mathrm{Tr}\bigl(\rho_A\log\rho_A\bigr) + \mathrm{Tr}\bigl(\rho_B\log\rho_B\bigr)
\;\equiv\; -S_A - S_B .
\end{aligned}
\end{equation}

\subsection{Efficient calculation of the 2-SRDMs}
In preparation for the construction of the 2-SRDMs, the SOP factor matrices are first transformed to the truncated NSPO basis:

\begin{equation}
\left(\tilde{\nu}_a^{(\kappa)}\right)_{\Upsilon}
\;=\; \sum_{i,j=1}^{N_\kappa}
   U^{(\kappa)\dagger}_{ij,\Upsilon}
   \left(\nu_a^{(\kappa)}\right)_{ij},
\qquad \Upsilon = 1, \ldots, M_\kappa \;\ll\; N_\kappa^{2} .
\end{equation}

\noindent
Again, the fact that many of the SOP factor matrices $\boldsymbol{\nu}_{a}^{(\kappa)}$ will be either diagonal or unit matrices can be leveraged. Using the transformed factor vectors $\boldsymbol{\tilde{\nu}}_{a}^{(\kappa)}$, the 2-SRDMs in the NSPO basis can be computed as

\begin{equation}\label{eq:srdm2-natsop}
R^{(\kappa,\lambda)}_{\Upsilon_\kappa \Upsilon_\lambda;\,\Upsilon'_\kappa \Upsilon'_\lambda}
\;=\; \sum_{a,b=1}^{S}
   \Gamma^{(\kappa,\lambda)}_{ab}\,
   \left(\tilde{\nu}_a^{(\kappa)}\right)_{\Upsilon_\kappa}
   \left(\tilde{\nu}_a^{(\lambda)}\right)_{\Upsilon_\lambda}
   \left(\tilde{\nu}_b^{(\kappa)}\right)^{\!*}_{\Upsilon'_\kappa}
   \left(\tilde{\nu}_b^{(\lambda)}\right)^{\!*}_{\Upsilon'_\lambda} .
\end{equation}

Introducing the intermediates

\begin{equation}
\Phi^{(b)}_{\Upsilon'_\kappa \Upsilon'_\lambda}
\;=\; \left(\tilde{\nu}_b^{(\kappa)}\right)^{\!*}_{\Upsilon'_\kappa}
       \left(\tilde{\nu}_b^{(\lambda)}\right)^{\!*}_{\Upsilon'_\lambda} .
\end{equation}

\noindent
and

\begin{equation}
\Lambda^{(a,b)}_{\Upsilon_\kappa \Upsilon_\lambda}
\;=\; \Gamma^{(\kappa,\lambda)}_{ab}\,
   \left(\tilde{\nu}_a^{(\kappa)}\right)_{\Upsilon_\kappa}
   \left(\tilde{\nu}_a^{(\lambda)}\right)_{\Upsilon_\lambda} .
\end{equation}

\noindent
the 2-SRDM $\boldsymbol{R}^{(\kappa,\lambda)}$ can be computed efficiently using the procedure detailed in Algorithm~\ref{alg:srdm2}.

\begin{algorithm}[H]
\caption{Construction of the 2-SRDM $R^{(\kappa,\lambda)}$ on the
truncated NSPO basis}
\label{alg:srdm2}
\begin{algorithmic}[1]
\Require Degree of freedom pair $(\kappa,\lambda)$ with $\kappa>\lambda$;
         retained NSPO dimensions $M_\kappa$, $M_\lambda$;
         natural-SOP factors
         $\bigl\{\tilde{\boldsymbol{v}}^{(\kappa)}_{a},\,
                 \tilde{\boldsymbol{v}}^{(\lambda)}_{a} | a=1,\dots,S\bigr\}$ and
         coefficients $\{\mu_a | a=1,\dots,S\}$;
         pre-computed Hilbert--Schmidt overlap tensor
         $\Omega \in \mathbb{C}^{d\times S\times S}$;
         pre-allocated buffers
         $R \in \mathbb{C}^{M_\kappa M_\lambda \times M_\kappa M_\lambda}$,
         $\Phi, \Lambda \in \mathbb{C}^{M_\kappa M_\lambda}$.
\State $R \gets 0$
\For{$b = 1, \ldots, S$}
    \For{$\Upsilon_{\lambda} = 1, \ldots, M_\lambda$}
        \For{$\Upsilon_{\kappa} = 1, \ldots, M_\kappa$}
        \State $\Phi_{\Upsilon_{\kappa} \Upsilon_{\lambda}} \gets
              \bigl(\tilde{\boldsymbol{v}}^{(\kappa)}_{b}\bigr)^{*}_{\Upsilon_{\kappa}}
              \bigl(\tilde{\boldsymbol{v}}^{(\lambda)}_{b}\bigr)^{*}_{\upsilon_{\lambda}}$
    \EndFor
  \EndFor
  \For{$a = 1, \ldots, S$}
    \State $\Gamma \gets \mu_a\,\mu_b^{*}\!\displaystyle\prod_{\gamma\neq\kappa,\lambda}\Omega^{(\gamma)}_{ba}$
    \For{$\Upsilon'_{\lambda} = 1, \ldots, M_\lambda$}
        \For{$\Upsilon'_{\kappa} = 1, \ldots, M_\kappa$}
        \State $\Lambda_{\Upsilon'_{\kappa} \Upsilon'_{\lambda}} \gets \Gamma\,
                \bigl(\tilde{\boldsymbol{v}}^{(\kappa)}_{a}\bigr)_{\Upsilon'_{\kappa}}\,
                \bigl(\tilde{\boldsymbol{v}}^{(\lambda)}_{a}\bigr)_{\Upsilon'_{\lambda}}$
      \EndFor
    \EndFor
    \State $R \gets R + \Phi \otimes \Lambda$
  \EndFor
\EndFor
\State \textbf{return} $R$
\end{algorithmic}
\end{algorithm}

\begin{algorithm}
\caption{Construction of the 2-SRDM $R^{(\kappa,\lambda)}$ on the truncated NSPO basis}
\begin{algorithmic}[1]
\Require Degree of freedom pair $(\kappa,\lambda)$ with $\kappa > \lambda$;
    retained NSPO dimensions $M_\kappa,\,M_\lambda$;
    natural-SOP factors $\{\tilde{\boldsymbol{\nu}}_a^{(\kappa)},\,\tilde{\boldsymbol{\nu}}_a^{(\lambda)}\,|\,a=1,\ldots,S\}$
    and coefficients $\{\mu_a\,|\,a=1,\ldots,S\}$;
    pre-computed Hilbert--Schmidt overlap tensor $\boldsymbol{\Omega}\in\mathbb{C}^{d\times S\times S}$;
    pre-allocated buffers $R\in\mathbb{C}^{M_\kappa M_\lambda\times M_\kappa M_\lambda}$,
    $\Lambda,\Phi\in\mathbb{C}^{M_\kappa M_\lambda}$.
\State $R \gets 0$
\For{$b = 1,\ldots,S$}
    \For{$\Upsilon'_\lambda = 1,\ldots,M_\lambda$}
        \For{$\Upsilon'_\kappa = 1,\ldots,M_\kappa$}
            \State $\Phi_{\Upsilon'_\kappa \Upsilon'_\lambda}
              \gets
              \bigl(\tilde{\nu}_b^{(\kappa)}\bigr)^{\!*}_{\Upsilon'_\kappa}
              \bigl(\tilde{\nu}_b^{(\lambda)}\bigr)^{\!*}_{\Upsilon'_\lambda}$
        \EndFor
    \EndFor
    \For{$a = 1,\ldots,S$}
        \State $\Gamma \gets \mu_a\,\mu_b^{*}\!\prod_{\gamma\neq\kappa,\lambda}\Omega^{(\gamma)}_{ba}$
        \For{$\Upsilon_\lambda = 1,\ldots,M_\lambda$}
            \For{$\Upsilon_\kappa = 1,\ldots,M_\kappa$}
                \State $\Lambda_{\Upsilon_\kappa \Upsilon_\lambda}
                  \gets
                  \Gamma\,
                  \bigl(\tilde{\nu}_a^{(\kappa)}\bigr)_{\Upsilon_\kappa}
                  \bigl(\tilde{\nu}_a^{(\lambda)}\bigr)_{\Upsilon_\lambda}$
            \EndFor
        \EndFor
        \State $R \gets R + \Phi \otimes \Lambda$
    \EndFor
\EndFor
\State \Return $R$
\end{algorithmic}
\end{algorithm}

\section{Maximum MPO bond dimension of Henon-Heiles Hamiltonian}

Using the finite-state automaton framework described in \cite{McCulloch_2007,PhysRevA.78.012356,Catarina2023}, we can identity the bond dimension of a MPO core by considering the possible ``virtual states ''. For the HH Hamiltonian given as

\begin{equation}
    H=\sum_i^N h_i + \lambda\sum_i^{N-1}q_k^2q_{k+1}-\lambda\sum_i^{N-1}q^3_{k+1}
\end{equation}

\begin{equation}
\hat{H} \;=\; \sum_{i=1}^{N} \hat{h}_i
            \;+\; \lambda \sum_{i=1}^{N-1} \hat{q}_i^{2}\,\hat{q}_{i+1}
            \;-\; \lambda \sum_{i=1}^{N-1} \hat{q}_{i+1}^{3} .
\end{equation}

The possible terms are thus the following:

\begin{equation}
    \begin{aligned}
        \cdots& \otimes^{3} \mathbb{I} \otimes^{3} \hat{h}_i \otimes^{1} \mathbb{I} \cdots \\
        \cdots& \otimes^{3} \mathbb{I} \otimes^{3} \lambda \hat{q}_i^{2} \otimes^{2} \hat{q}_{i+1} \otimes^{1} \mathbb{I} \cdots \\
        \cdots& \otimes^{3} \mathbb{I} \otimes^{3} \mathbb{I} \otimes^{3} \bigl(-\lambda \hat{q}_i^{3}\bigr) \otimes^{I} \mathbb{1} \cdots
\end{aligned}
\end{equation}

\noindent where the superscript on top of the tensor products denote the possible ``virtual states'' in the automaton.

\begin{itemize}
    \item State 1: Only Identity to the right
    \item State 2: One $\hat{q}_{k+1}$ term to the right, followed by identity
    \item State 3: One complete term to the right
\end{itemize}

The three number of states would then translate to a bulk MPO core tensor with bond dimensions of three, in this case constant through the entire MPO chain. The individual bulk core tensor of the MPO $W^{[k]}\in\mathbb{C}^{w_{k-1}\times w_k}$ can be expressed as

\begin{equation}
    W^{[k]}=\begin{pmatrix}
        \mathbb{I} & 0 & 0 \\
        \hat{q}_{i} & 0 & 0 \\
        \hat{h}_i-\lambda\hat{q}^3_{i+1}& \lambda\hat{q}_{i}^2 & \mathbb{I}
    \end{pmatrix}
\end{equation}

\noindent where each element is a $N_k\times N_k$ matrix.

Similarly, we can then see that if the operator has a product form such as $\hat{O}=\hat{o}_1\otimes\cdots\otimes\hat{o}_d$, then the only possible virtual state would be a complete term to the right, thus corresponding to a bond dimension of one.

\section{Sub-additivity}
Klein's inequality theorem states that for any $n\times n$ Hermitian matrices $A,B$ and a convex function $f$, the following inequality is satisfied\cite{Carlen2009TRACEIA, BHATIA2003125}:

\begin{equation}\label{eq:klein_inequality}
    \tr{(f(A)-f(B)-(A-B)f'(B))}\geq 0.
\end{equation}

Taking the function to be the vNEE, $f(x)=x\log(x)$, which is convex in $x$, and substituting $A\equiv\rho_{AB}$ and $B\equiv\rho_{A}\otimes\rho_B$ into Eq.~\eqref{eq:klein_inequality} sets up the inequality

\begin{equation}\label{eq:subadditivity_deriv_1}
\begin{aligned}
    \tr{(\rho_{AB}\log(\rho_{AB}))}-\tr{(\rho_{AB}\log(\rho_{A}\otimes\rho_{B}))} &\geq \tr{(\rho_{AB})}-\tr{(\rho_{A}\otimes\rho_{B})}, \\
    \tr{(\rho_{AB}\log(\rho_{AB}))}-\tr{(\rho_{AB}\log(\rho_{A}\otimes\rho_{B}))} &\geq 0 \quad.
\end{aligned}
\end{equation}

\noindent
Note that, in the second line of Eq.~\eqref{eq:subadditivity_deriv_1}, the right hand side is zero because: (i) $\tr{(A \otimes B)} = \tr{(A)} \tr{(B)}$ for square matrices $A,B$, and; (ii) any partition of the density $\rho$ has unit trace. Lastly, using the properties\cite{BHATIA2003125}

\begin{subequations}
\label{eq:subadditivity_deriv_property}
    \begin{align}
       \log(\rho_{A}\otimes\rho_{B})&=\log(\rho_{A})\otimes\mathbb{I}+\mathbb{I}\otimes\log(\rho_{B}) \\
        \tr{(\rho_{AB}[X\otimes\mathbb{I}])} &= \tr{(\rho_AX)},
    \end{align}
\end{subequations}

\noindent for all Hermitian matrix $X$, the second term in the second line of Eq.~\eqref{eq:subadditivity_deriv_1} simplifies to


\begin{equation}\label{eq:subadditivity_deriv_2}
    \begin{aligned}
        \tr{(\rho_{AB}\log(\rho_{A}\otimes\rho_{B}))} &= \tr{(\rho_{AB}[\log(\rho_A)\otimes\mathbb{I}+\mathbb{I}\otimes\log(\rho_B)])} \\
        &= \tr{(\rho_{AB}[\log(\rho_A)\otimes\mathbb{I}])} + \tr{(\rho_{AB}[\mathbb{I}\otimes\log(\rho_B)])}\\
        &=\tr{(\rho_{A}\log\rho_A)}+\tr{(\rho_{B}\log\rho_B)} \equiv -S_A-S_B
    \end{aligned}    
\end{equation}

It is important to note that Eq.~\eqref{eq:subadditivity_deriv_property} does not require additional constraints on the class of the operators, in this case $\rho_{A(B)}$, except that it admits a spectral decomposition, which is satisfied by any Hilbert-Schmidt operator since they are compact\cite{Sołtan2018}, as given by the relation in Eq.~\eqref{eq:operator_class_relation} in the main text. Combining Eq.~\eqref{eq:subadditivity_deriv_2} and Eq.~\eqref{eq:subadditivity_deriv_1}, we obtain the sub-additivity property $S_{A}+S_{B}\geq S_{AB}$. Since unit trace implies that $\tr{(A)}=1 \Rightarrow \sum_i \lambda_i = 1$, where $\{\lambda_i\}$ are the eigenvalues of $A$, the eigenvalues of the 1- and 2-SRDMs are normalized to have unit trace. Consequently, the subadditivity and the inequality in Eq.~\eqref{eq:mi} continue to hold for the Hilbert-Schmidt operators used in this work.

\bibliography{opt}